\documentclass[12pt]{article}
\usepackage{hyperref}
\usepackage[english]{babel}
\usepackage{amsmath}
\usepackage{latexsym}
\usepackage{amssymb}
\usepackage[dvips]{graphicx}
\usepackage{epsfig}
\usepackage{psfrag}

\numberwithin{equation}{section} \numberwithin{table}{section}
\numberwithin{figure}{section}

\begin{document}



\begin{titlepage}
   \begin{flushright}
{\small MPP-2013-285 }
  \end{flushright}

   \begin{center}

     \vspace{20mm}

     {\LARGE \bf Holographic entanglement entropy of semi-local quantum liquids

     \vspace{3mm}}

     \vspace{10mm}

    Johanna Erdmenger, Da-Wei Pang and Hansj\"{o}rg Zeller

     \vspace{5mm}
      {\small \sl Max-Planck-Institut f\"{u}r Physik (Werner-Heisenberg-Institut)\\
      F\"{o}hringer Ring 6, 80805 M\"{u}nchen, Germany}\\

     {\small \tt jke,dwpang, zeller@mppmu.mpg.de}
     \vspace{10mm}

   \end{center}

\begin{abstract}
\baselineskip=18pt
We consider the holographic entanglement entropy of $(d+2)$-dimensional semi-local quantum liquids, for which the dual gravity background in the deep interior is $AdS_{2}\times\mathbb{R}^{d}$ multiplied by a warp factor which depends on the radial coordinate. The entropy density of this geometry goes to zero in the extremal limit. The thermodynamics associated with this semi-local background is discussed via dimensional analysis and scaling arguments. For the case of an asymptotically AdS UV completion of this geometry, we show that the entanglement entropy of a strip and an annulus exhibits a phase transition as a typical length of the different shapes is varied, while there is no sign of such a transition for the entanglement entropy of a sphere. Moreover, for the spherical entangling region, the leading order contribution to the entanglement entropy in the IR is calculated analytically. It exhibits an area law behaviour and agrees with the numerical result.
\end{abstract}
\setcounter{page}{0}
\end{titlepage}

\pagestyle{plain} \baselineskip=19pt

\tableofcontents

\section{Introduction}
The AdS/CFT correspondence~\cite{Maldacena:1997re, Aharony:1999ti}, or more generally, gauge/gravity duality,
has been proven to be a powerful tool for studying the dynamics of strongly coupled field theories.
This paradigm has been applied to understand the low-temperature physics of strongly-coupled electron systems (AdS/CMT),
such as superconductors~\cite{Hartnoll:2008vx, Hartnoll:2008kx} and (non-)Fermi liquids~\cite{Liu:2009dm, Cubrovic:2009ye, Faulkner:2009wj}.

In most of the realistic condensed matter systems, one basic ingredient is the presence of a finite charge density. Therefore we need a conserved global charge in the gravity dual, i.e.~we consider charged black hole solutions. The initial study of holographic systems at finite density focused on Reissner-Nordstr\"{o}m black holes in AdS space, which may be considered as the simplest laboratory for exploring AdS/CMT. In particular, the fermionic two-point function in this background displays the behavior of fermionic quasi-particles corresponding to a non-Fermi liquid. This is due to the emergent $AdS_{2}$ near-horizon geometry in the extremal RN-AdS background~\cite{Liu:2009dm, Cubrovic:2009ye, Faulkner:2009wj}. However, the RN-AdS black hole has a significant disadvantage from a condensed matter point of view: it has finite entropy at extremality, i.e.~at zero temperature.

A further step towards constructing gravity duals of strongly-coupled systems at finite density is to include the leading order relevant scalar operators, which on the gravity side corresponds to the Einstein-Maxwell-Dilaton system with a scalar potential. This makes the theory flow to an IR fixed point which is not the near-horizon RN-AdS geometry. There are models in this class which have zero entropy at extremality and are therefore of interest for condensed matter applications. These models have been extensively studied in~\cite{Charmousis:2010zz}, where they are characterized by studying the thermodynamics, spectra and conductivities. The analysis is based on the concept of Effective Holographic Theory (EHT). The central point of EHT is to truncate a string theory to a finite spectrum of
low-lying states. Intuitively, we may argue that the truncation is reasonable if neglected states cannot become relevant in the UV or irrelevant in the IR. In~\cite{Charmousis:2010zz} the EHTs of
the Einstein-Maxwell-Dilaton theory were parametrized in terms of the IR asymptotics
of the scalar functions: the scalar potential and the nontrivial Maxwell coupling. Hence
the exact solutions of the Einstein-Maxwell-Dilaton theory describe the IR asymptotic
geometry. The EHT has the advantage that it provides descriptions of large
classes of IR dynamics, although the understanding of the dual field theory is less clear.

The $(d+2)$-dimensional Einstein-Maxwell-Dilaton theory admits the hyperscaling violation metric as an exact solution,
\begin{equation}
\label{hvmetric}
ds^{2}=\frac{1}{r^{2}}\left(-\frac{dt^{2}}{r^{2d(z-1)/(d-\theta)}}+r^{2\theta/(d-\theta)}dr^{2}+\sum\limits^{d}_{i=1}dx_{i}^{2}\right),
\end{equation}
where $z$ denotes the dynamical exponent and $\theta$ is the hyperscaling violation parameter. The background possesses the following scaling property,
\begin{equation}
t\rightarrow\lambda^{z}t,~~x_{i}\rightarrow\lambda x_{i},~~ds\rightarrow\lambda^{\theta/d}ds.
\end{equation}
The entropy density at finite temperature scales as $s\sim T^{(d-\theta)/z}$. It has been observed in~\cite{Hartnoll:2012wm} that for general finite $z$ and $\theta$, the behavior of the spectral densities in these spacetimes seem to better describe the properties of theories with bosonic degrees of freedom rather than with fermionic ones. However, in the same paper the authors consider the limit $z\rightarrow\infty$, which allows for low-energy modes at all momenta, resembling features found in fermionic systems. Furthermore, to avoid the undesirable ground state entropy density, we may take the following limits~\cite{Hartnoll:2012wm},
\begin{equation}
\label{smlimit}
z\rightarrow\infty,~~\theta\rightarrow-\infty,~~\eta\equiv-\frac{\theta}{z}~{\rm fixed}.
\end{equation}
Then the metric becomes
\begin{equation}
\label{smmetric}
ds^{2}=\frac{1}{r^{2}}\left(-\frac{dt^{2}}{r^{2d/\eta}}+\frac{dr^{2}}{r^{2}}+\sum\limits^{d}_{i=1}dx_{i}^{2}\right).
\end{equation}
This metric is conformal to $AdS_{2}\times\mathbb{R}^{d}$, which can be seen by taking a new radial coordinate $r=\xi^{\eta/d}$,
\begin{equation}
ds^{2}=\frac{1}{\xi^{\frac{2\eta}{d}}}\left[-\frac{dt^{2}}{\xi^{2}}+\frac{d\xi^{2}}{\xi^{2}}+\sum\limits^{d}_{i=1}dx_{i}^{2}\right].
\end{equation}
In the corresponding non-extremal solution the entropy density scales as $s\sim T^{\eta}$, which means that the entropy density goes to zero in the extremal limit. In~\cite{Iqbal:2011in} the $AdS_{2}\times\mathbb{R}^{d}$ near-horizon geometry of the $(d+2)$-dimensional extremal RN-AdS black hole is referred to as a {\it holographic semi-local quantum liquid}, characterized by a finite spatial correlation length, an infinite correlation time and a non trivial scaling behavior in the time direction. Since the background~(\ref{smmetric}) is conformal to $AdS_{2}\times\mathbb{R}^{d}$, it may be seen as a generalization of the dual of holographic semi-local quantum liquids. Backgrounds with hyperscaling violation and semi-locality were first investigated in~\cite{Gouteraux:2011ce}.

It is straightforward to characterize properties of Fermi surfaces in backgrounds with semi-locality by performing an analysis of the fermionic correlations as in~\cite{Faulkner:2009wj}. However, there is a further quantity which may help in characterizing the presence of Fermi surfaces: the entanglement entropy. It was conjectured in~\cite{Ogawa:2011bz} that systems with Fermi surfaces exhibit a logarithmic violation of the `area law' behavior of the entanglement entropy. In the same paper, the authors construct a gravity dual which displays the expected behavior for non-Fermi liquids. Furthermore, it has been shown that when the hyperscaling violation parameter $\theta=d-1$, the background~(\ref{hvmetric}) also exhibits a violation of the `area law'~\cite{Huijse:2011ef}. However, the interpretation of this violation as a sign for a Fermi surface stands in contrast to the results of~\cite{Hartnoll:2012wm}, where the spectral density does not seem to describe a fermionic system (see discussion above eq.~\eqref{smlimit}). For subsequent developments in this direction, see~\cite{others}.

\begin{figure}
\begin{center}
\includegraphics[angle=0,width=0.6\textwidth]{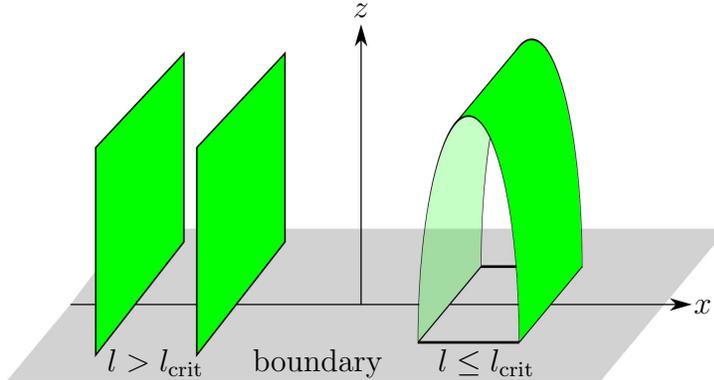}\put(-142,138){$z$}\put(-13,26){$x$}\put(-180,4){boundary}\put(-235,4){$l>l_{\rm crit}$}\put(-110,4){$l\leq l_{\rm crit}$}
\caption{Connected and disconnected solution for the strip case.}
\label{fig:strip}
\end{center}
\end{figure}

One may wonder how the entanglement entropy will behave if we take the limit~(\ref{smlimit}) in the background with hyperscaling violation.
It has been observed in~\cite{Hartnoll:2012wm} that when the entangling region on the boundary is a strip, then only for a strip width $l=l_{\rm crit}$ there is a connected minimal surface solution (see figure~\ref{fig:strip}). For all other values of $l$ the solution is a disconnected minimal surface, i.e.~two slabs reaching into the bulk without ever touching each other. It was conjectured in~\cite{Hartnoll:2012wm} that if the hyperscaling violating geometry is an IR completion of an asymptotically AdS spacetime, then the connected minimal surface may exist for separation lengths $l<l_{\rm crit}$, while for $l>l_{\rm crit}$ two disconnected minimal surfaces dominate. This describes a phase transition between the disconnected and connected solutions. Interestingly, a similar behavior is observed in confining geometries~\cite{Klebanov:2007ws}. The holographic entanglement entropy of five-dimensional extremal two-charge black hole in type IIB supergravity was considered in~\cite{Kulaxizi:2012gy}, where the near-horizon geometry is of the type~(\ref{smmetric}) with $d=3, \eta=1$. Also in this case, the same behavior of the entanglement entropy for the strip as described above was found. An advantage of the background studied in~\cite{Kulaxizi:2012gy} is that the full geometry is explicitly known, hence the following picture may emerge: For sufficiently large boundary separation length $l$ the hypersurface in the bulk should probe the IR limit of the geometry, which means that the background may be approximated by the semi-local geometry. Then there exists a maximal value of $l=l_{\rm crit}$ beyond which only the disconnected hypersurfaces contribute. For sufficiently small $l$, the entangling surface probes the UV and the full geometry should be taken into account. These arguments are confirmed by numerics in~\cite{Kulaxizi:2012gy}, where the authors also state that the transition at $l_\text{crit}$ is second order. On the other hand, for a spherical entangling region a phase transition of this type was not observed.

In this paper we study holographic entanglement entropy of $(d+2)$-dimensional semi-local quantum liquids for general $\eta$. For completeness we first review the exact solution with semi-locality both at extremality and at finite temperature. Even though the solutions only describe the IR geometry, we may still study their thermodynamical properties by dimensional analysis and scaling arguments. Then we calculate the holographic entanglement entropy in the extremal background with the entangling surfaces being a strip and a sphere. For the strip case we find, similarly to the cases discussed above, that there exists only a connected solution if the boundary separation length $l$ is constant. For the sphere case, we are able to calculate the entanglement entropy analytically and find that the leading order contribution exhibits an area law behavior. As discussed in the previous paragraph, the full geometry is needed if the boundary separation length is sufficiently small, therefore we construct the full $(d+2)$ dimensional geometry for generic values of $\eta$ (see eq.~\eqref{smlimit}), which is asymptotically AdS and possesses semi-locality in the IR. We compute the holographic entanglement entropy in this geometry. For the strip case, the behavior of the entanglement entropy is as expected: the connected hypersurface dominates when the boundary separation length $l$ is small, while the disconnected hypersurfaces dominate when $l>l_{\rm crit}$. However, in the sphere case we do not find such a transition. Finally as proposed in~\cite{Kulaxizi:2012gy}, we calculate the entanglement entropy for an annulus entanglement surface in order to interpolate between the sphere and the strip case. The annulus is supposed to approximate the spherical entanglement entropy behavior when the inner radius is very small compared to the outer radius, while the same behavior as in the strip case is obtained for both radii large and their difference small. We find that there is a transition taking place between two concentric spheres (disconnected solution) and a deformed annulus (connected solution) at a critical value $(\Delta \rho)_\text{crit}$ of the difference between the outer and inner radius. Several aspects of this transition are very interesting: First as opposed to the strip case where the transition from the disconnected to the connected solution is second order, here, depending on the dimension and the value of the inner and outer radii we find a swallow tail behavior, known from first order phase transitions. For larger radii we see a second order transition, this is an indication that for large radii we are indeed approximating the strip case. Second, the maximal radii difference $(\Delta \rho)_\text{max}$ for which a connected solution exists, approximates $l_\text{crit}$ (critical width of the strip) with increasing values of the radii. Note that $(\Delta \rho)_\text{max}=(\Delta \rho)_\text{crit}$ only in the cases where we find a second order transition. Finally, we do not find a solution with vanishing inner radius in order to approximate the sphere. This is due to the fact that for decreasing values of the outer radius, the difference between the radii also decreases, with the difference being smaller.

The paper is organized as follows: We review the exact solutions both at extremality and finite temperature and study the corresponding thermodynamics in section 2. Then we calculate the entanglement entropy in the extremal background for both the strip and the sphere cases in section 3. After constructing solutions asymptotic to AdS in the UV in section 4, we revisit the holographic entanglement entropy for entangling regions being a strip, a sphere and an annulus in section 5. A summary and an interpretation of the results are given in section 6.
\section{The background with semi-locality}
\label{sec:back}
In this section we study the background with semi-locality. After reviewing the solutions both at extremality and at finite temperature, we will study the thermodynamics by dimensional analysis and scaling arguments.

\subsection{The background}
We start from the action of Einstein-Maxwell-Dilaton theory,
\begin{equation}
 S=\int d^{d+2}x\sqrt{-g}\left(\frac{1}{2\kappa^{2}}R-\frac{Z(\Phi)}{4e^{2}}F_{\mu\nu}F^{\mu\nu}
 -\frac{1}{\kappa^{2}}(\partial\Phi)^{2}-\frac{1}{2\kappa^{2}L^{2}}V(\Phi)\right),
\end{equation}
with effective gauge coupling and scalar potential
\begin{equation}
 Z(\Phi)=Z_{0}^{2}e^{\alpha\Phi},~~~V(\Phi)=-V_{0}^{2}e^{-\beta\Phi}.
\end{equation}
Here $Z_{0}, V_{0}, \alpha, \beta$ are constants characterizing the theory. Theories of this type were named ``Effective Holographic Theory''
in~\cite{Charmousis:2010zz}. The backgrounds with hyperscaling violation and general semi-locality as used in the subsequent sections were first investigated in~\cite{Gouteraux:2011ce}.
The equations of motion are given by
\begin{eqnarray}
 & &\partial_{\mu}(\sqrt{-g}Z(\Phi)F^{\mu\nu})=0,\nonumber\\
 & &\partial_{\mu}(\sqrt{-g}\partial^{\mu}\Phi)=\frac{\kappa^{2}}{8e^{2}}\sqrt{-g}\frac{\partial Z}{\partial\Phi}F_{\rho\sigma}F^{\rho\sigma}
 +\frac{1}{4L^{2}}\sqrt{-g}\frac{\partial V}{\partial\Phi},\nonumber\\
 & &R_{\mu\nu}-\frac{1}{2}Rg_{\mu\nu}-2\partial_{\mu}\Phi\partial_{\nu}\Phi+g_{\mu\nu}(\partial\Phi)^{2}\nonumber\\
 & &-\frac{\kappa^{2}}{e^{2}}Z(\Phi)F_{\mu\lambda}{F_{\nu}}^{\lambda}+\frac{\kappa^{2}}{4e^{2}}Z(\Phi)g_{\mu\nu}F_{\rho\sigma}F^{\rho\sigma}
 +\frac{V(\Phi)}{2L^{2}}=0.
\end{eqnarray}

It was observed in~\cite{Huijse:2011ef} that the above theory admits the exact solution
\begin{eqnarray}
 & &ds^{2}=\frac{L^{2}}{r^{2}}\left(-f(r)dt^{2}+g(r)dr^{2}+\sum\limits^{d}_{i=1}dx_{i}^{2}\right),\nonumber\\
 & &f(r)=f_{0}r^{-\frac{2d(z-1)}{d-\theta}},~~~g(r)=g_{0}r^{\frac{2\theta}{d-\theta}},
\end{eqnarray}
where $f_{0}$ and $g_{0}$ are constants determined by $Z(\Phi)$ and $V(\Phi)$, which will not be explicitly
written down here. $\theta$ is the hyperscaling violation parameter and $z$ is the dynamical exponent, which are determined
by $\alpha$ and $\beta$,
\begin{equation}
\label{thetaz}
 \theta=\frac{d^{2}\beta}{\alpha+(d-1)\beta},~~~z=1+\frac{\theta}{d}+\frac{8(d(d-\theta)+\theta)^{2}}{d^{2}(d-\theta)\alpha^{2}}.
\end{equation}
We are interested in the limit
\begin{equation}
\label{limit}
z\rightarrow\infty, ~~\theta\rightarrow-\infty ~~{\rm while} ~~\eta\equiv-\theta/z ~~{\rm fixed},
\end{equation}
following~\cite{Hartnoll:2012wm}. This
requirement leads to
\begin{equation}
 \beta=-\frac{\sqrt{8/d}}{1+d/\eta},~~~\alpha=-(d-1)\beta,
\end{equation}
which can be easily obtained by taking such a limit in~(\ref{thetaz}). Then
the solution at extremality is given by
\begin{eqnarray}
\label{extIR}
 & &ds^{2}_{d+2}=\frac{L^{2}}{r^{2}}\left(-\frac{dt^{2}}{r^{2d/\eta}}+\frac{g_{0}}{r^{2}}dr^{2}+\sum\limits^{d}_{i=1}dx_{i}^{2}\right),\nonumber\\
 & &g_{0}=\frac{d^{2}}{V_{0}^{2}}\left(1+\frac{1}{\eta}\right)^{2},~~\Phi=\sqrt{\frac{d}{2}}\sqrt{1+\frac{d}{\eta}}\log r,\nonumber\\
 & &A_{t}=\frac{eL}{\kappa}h(r),~~~h(r)=\frac{h_{0}}{r^{d(1+1/\eta)}},~~~h_{0}=\frac{1}{Z_{0}\sqrt{1+\eta}}.
\end{eqnarray}
Such a background possesses the following scaling properties
\begin{equation}
 t\rightarrow\lambda t, ~~~r\rightarrow\lambda^{\eta/d}r,~\Rightarrow~ds\rightarrow\lambda^{-\eta/d}ds.
\end{equation}
 Furthermore, in this background
only $t$ and $r$ are involved in the scaling symmetries while the spatial coordinates $x_{i}$ are spectators, hence the background geometry is ``semi-local''~\cite{Iqbal:2011in} and it can be easily seen that it is conformal to $AdS_{2}\times\mathbb{R}^{d}$.

The finite-temperature counterparts can be written as follows
\begin{equation}
\label{bh}
 ds^{2}_{d+2}=\frac{L^{2}}{r^{2}}\left(-\frac{\chi(r)dt^{2}}{r^{2d/\eta}}+\frac{g_{0}}{r^{2}\chi(r)}dr^{2}+\sum\limits^{d}_{i=1}dx_{i}^{2}\right),
 ~~\chi(r)=1-\left(\frac{r}{r_{h}}\right)^{d(1+1/\eta)},
 \end{equation}
 while the other field configurations remain invariant as in the extremal case. The temperature and entropy density of this black hole are given by
 \begin{equation}
  T=\frac{V_{0}}{4\pi}r_{h}^{-d/\eta},~~~s=\frac{L^{d}}{4r_{h}^{d}}.
 \end{equation}
Note that we always have $s\sim T^{\eta}$, irrespective of the number of spatial dimensions.
\subsection{Thermodynamics}
Let us study the thermodynamics of the semi-local geometry. Generally, the full solution should be considered when considering the thermodynamics, while for our case the exact solution just describes the IR geometry. However, we can still discuss the thermodynamics by dimensional analysis and scaling arguments, following~\cite{Goldstein:2009cv, Chen:2010kn}.

As discussed in previous subsection, it can easily be obtained that
\begin{equation}
T\propto r_{h}^{-d/\eta},~~~s\propto r_{h}^{-d},~~\Rightarrow~~s\sim T^{\eta},
\end{equation}
which holds in arbitrary dimensions. Note that the scaling dimensions of the temperature $T$ and the chemical potential $\mu$ are both of $[{\rm Mass}]^{-1}$, so the entropy density scales as $s\sim T^{\eta}\mu^{d-\eta}$. On the other hand, in $(d+2)-$dimensional bulk spacetime, the entropy may be evaluated from the on-shell action. Therefore a prefactor $L^{d}/G_{N}$ should exist, where $G_{N}$ denotes the Newton's constant. Thus we have
\begin{equation}
s=aCT^{\eta}\mu^{d-\eta},
\end{equation}
where $C\sim L^{d}{G_{N}}$ and $a$ depends on the coupling constant $\eta$. The specific heat is given by
\begin{equation}
C_{V}=T\left(\frac{ds}{dT}\right)_{\mu}=aC\eta T^{\eta}\mu^{d-\eta},
\end{equation}
which is always positive. The other thermodynamical quantities are determined by the Gibbs-Duhem relation
$$sdT+nd\mu-dP=0,$$
where $P$ is the pressure and $n$ denotes the number density.

The pressure reads
\begin{equation}
P=\frac{a}{\eta+1}CT^{\eta+1}\mu^{d-\eta}+bCe^{(d+1)\eta\phi_{0}}\mu^{d+1},
\end{equation}
where the first term can be obtained by integrating the Gibbs-Duhem relation while keeping $\mu$ fixed, and the second term can be fixed by dimensional analysis. Here $\phi_{0}$ is the asymptotic value of the dilaton. The number density is given by
\begin{equation}
n=\frac{\partial P}{\partial\mu}=\frac{a(d-\eta)}{\eta+1}CT^{\eta+1}\mu^{d-\eta-1}+b(d+1)Ce^{(d+1)\eta\phi_{0}}\mu^{d}.
\end{equation}
Finally the energy density is
\begin{equation}
\rho=Ts+\mu n-P=\frac{d}{\eta+1}aCT^{\eta+1}\mu^{d-1}+bdCe^{(d+1)\eta\phi_{0}}\mu^{d+1},
\end{equation}
which leads to
\begin{equation}
P=\frac{1}{d}\rho.
\end{equation}
Note that the results are valid when $T\ll\mu$. As the temperature increases for fixed $\mu$, the geometry is no longer a good approximation and the corrections to the above formulae will become important.
Moreover, the suscetibility is given by
\begin{equation}
\chi=\left(\frac{\partial n}{\partial\mu}\right)_{T}=\frac{(d-\eta)(d-\eta-1)}{\eta+1}aCT^{\eta+1}\mu^{d-\eta-2}+bd(d+1)Ce^{(d+1)\eta\phi_{0}}\mu^{d-1}.
\end{equation}
Note that when $d>\eta+1$ or $d<\eta$, the first term is positive, when $\eta<d<d+1$, the first term is negative, while the second term is always positive.  The susceptibility characterizes the stability of the system. However, to determine whether $\chi$ actually turns negative, which signals a phase transition, requires to consider the regime beyond the limit $T\ll\mu$.
\section{Holographic entanglement entropy in the semi-local background}
In this section we calculate the holographic entanglement entropy in the background with semi-locality, with the entangling surface being a strip and a sphere. For this purpose we consider the case where $g_{0}=1$ in~(\ref{extIR}),
\begin{equation}
\label{simple}
ds^{2}=\frac{L^{2}}{z^{2}}\left[-\frac{dt^{2}}{z^{p}}+\frac{dz^{2}}{z^{2}}+\sum\limits^{d}_{i=1}dx_{i}^{2}\right],~~p=2d/\eta.
\end{equation}
This metric is conformal to $AdS_{2}\times\mathbb{R}^{d}$, which may be seen explicitly after taking the coordinate transformation $z=\xi^{2/p}$,
\begin{equation}
\label{confads2}
ds^{2}=\frac{L^{2}}{\xi^{\frac{2\eta}{d}}}\left[-\frac{dt^{2}}{\xi^{2}}+\frac{d\xi^{2}}{\xi^{2}}+\sum\limits^{d}_{i=1}dx_{i}^{2}\right].
\end{equation}
The metric~(\ref{simple}) is used when calculating the entanglement entropy of a strip, while the metric~(\ref{confads2}) is considered when dealing with the case of a sphere. For the strip case we find that the boundary separation length is always constant, which means that in the deep IR, the disconnected surfaces dominate. For the sphere case we are able to extract the leading order behavior of the entanglement entropy analytically, following~\cite{andrei}\footnote{The corresponding subsection is based on unpublished notes~\cite{andrei} for $d=2,~3$ cases were studied in, which we generalize to arbitrary $d$.}. The analytic results are confirmed by numerical evaluations and the leading order behavior exhibits an 'area law'.
\subsection{The strip}
The holographic entanglement entropy (HEE) in Einstein gravity is determined by~\cite{Ryu:2006bv, Ryu:2006ef, Nishioka:2009un}
\begin{equation}
S_{A}=\frac{\rm Area(\gamma_{A})}{4G_{N}},
\end{equation}
where $G_{N}$ denotes the Newton constant and $\gamma_{A}$ is the codimension two minimal area surface which coincides with $\partial A$ at the boundary.
This formula has been proven in~\cite{Casini:2011kv} for a spherical entangling region and in~\cite{Lewkowycz:2013nqa} for more general cases. Let us consider the strip case,
\begin{equation}
x_{1}\equiv x\in\left[-\frac{l}{2},\frac{l}{2}\right],~~x_{i}\in[0,L_{x}],~~i=2,\cdots,d,
\end{equation}
where $l\ll L_{x}$. The induced metric is given by
\begin{equation}
ds^{2}_{\rm ind}=\frac{L^{2}}{z^{2}}\left(\left(\frac{1}{z^{2}}+x^{\prime2}\right)dz^{2}+\sum\limits^{d}_{i=1}dx_{i}^{2}\right),
\end{equation}
where we have parameterized the minimal surface area $\gamma_{A}$ by $x=x(z)$. Therefore the minimal surface area reads
\begin{eqnarray}
A(\gamma)&=&2\int\frac{L^{d}}{z^{d}}\sqrt{\frac{1}{z^{2}}+x^{\prime2}}\nonumber\\
&=&2L^{d}L_{x}^{d-1}\int\frac{dz}{z^{d}}\sqrt{\frac{1}{z^{2}}+x^{\prime2}}.
\end{eqnarray}
Since the Lagrangian does not explicitly contain $x$, there exists a conserved quantity
\begin{equation}
C\equiv\frac{x^{\prime}}{z^{d}\sqrt{\frac{1}{z^{2}}+x^{\prime2}}},
\end{equation}
which leads to
\begin{equation}
x^{\prime}=\frac{(\frac{z}{z_{\ast}})^{d}}{z\sqrt{1-(\frac{z}{z_{\ast}})^{2d}}}.
\end{equation}
Here $z_{\ast}$ denotes the turning point where $x^{\prime}$ diverges.
The boundary separation length $l$ is related to $z_{\ast}$ by
\begin{equation}
\frac{l}{2}=\int^{z_{\ast}}_{0}dz\frac{(\frac{z}{z_{\ast}})^{d}}{z\sqrt{1-(\frac{z}{z_{\ast}})^{2d}}},
\end{equation}
which gives
\begin{equation}
l=l_{\rm crit}=\frac{\pi}{d},
\end{equation}
which is constant in arbitrary $d$ dimensions.

The constant boundary separation length has been observed for several other examples, for example, for NS5-branes in~\cite{Ryu:2006ef} and for backgrounds with semi-locality~\cite{Hartnoll:2012wm, Kulaxizi:2012gy}. As argued in~\cite{Hartnoll:2012wm}, this result indicates that a minimal surface connecting the lines at the boundary only exists for a specific separation $l=l_{\rm crit}$, and a connected minimal surface only exists for separations $l<l_{\rm crit}$. As $l\rightarrow l_{\rm crit}$, the minimal surface droops increasingly further into the IR and when $l>l_{\rm crit}$, the disconnected minimal surface ( two surfaces falling into the IR at constant separation) dominates. This behavior is reminiscent of holographic entanglement entropy in confined phases~\cite{Klebanov:2007ws}. Moreover, as claimed in~\cite{Kulaxizi:2012gy}, when $l$ is sufficiently large, the hypersurface should probe the IR geometry, which is just our background~(\ref{simple}). In this case there exists a maximal value $l=l_{\rm crit}$ which corresponds to the curved solution. The trivial solution $x^{\prime}=0$, i.e. disconnected hypersurface dominates when $l>l_{\rm crit}$. When $l$ is sufficiently small, the entangling surface should probe the UV of the geometry, and $l$ is expected to be a smooth function of $z_{\ast}$. We will see that this picture holds when working with the UV-completed geometry.
\subsection{The sphere}
In this subsection we calculate the holographic entanglement entropy with a spherical entangling surface.
For convenience we work with the metric~(\ref{confads2}), which is explicitly conformal to $AdS_{2}\times\mathbb{R}^{d}$.
The spherical entangling region is parameterized by
$$\sum\limits^{d}_{i=1}x_{i}^{2}=R^{2},$$
and the induced metric is given by
\begin{equation}
ds^{2}=\frac{L^{2}}{\xi^{\frac{2\eta}{d}}}\left[(1+\frac{\xi^{\prime2}}{\xi^{2}})d\rho^{2}+\rho^{2}d\Omega_{d-1}^{2}\right].
\end{equation}
We find that the minimal surface area reads
\begin{eqnarray}
\label{Agm}
A(\gamma)&=&L^{d}\int d\Omega_{d-1}d\rho\frac{\rho^{d-1}}{\xi^{\eta}}\sqrt{1+\frac{\xi^{\prime2}}{\xi^{2}}}\nonumber\\
&=&L^{d}{\rm Vol}(\Omega_{d-1})\int d\rho\frac{\rho^{d-1}}{\xi^{\eta}}\sqrt{1+\frac{\xi^{\prime2}}{\xi^{2}}},
\end{eqnarray}
which leads to the equation of motion
\begin{equation}
\label{eomsphere}
\frac{\partial}{\partial\rho}\left(\frac{\rho^{d-1}\xi^{\prime}}{\xi^{\eta+2}\sqrt{1+\frac{\xi^{\prime2}}{\xi^{2}}}}\right)=
-\frac{\rho^{d-1}}{\xi^{\eta+3}\sqrt{1+\frac{\xi^{\prime2}}{\xi^{2}}}}(\eta\xi^{2}+(\eta+1)\xi^{\prime2}).
\end{equation}
\begin{figure}
\label{xirho}
\begin{center}
\vspace{-1cm}
\hspace{-0.5cm}
\includegraphics[angle=0,width=0.45\textwidth]{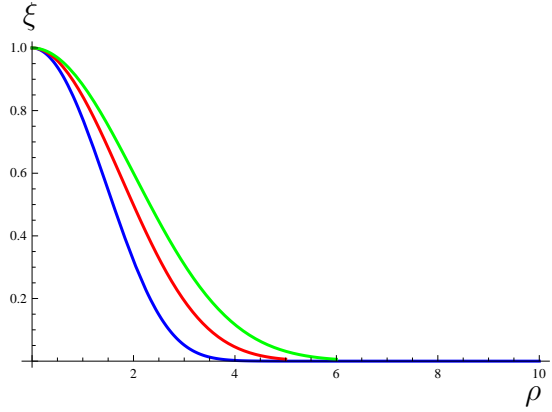}\put(-200,135){$\xi$}\put(-10,-7){$\rho$}
\caption{\small The plot of the embedding profile $\xi(\rho)$, which is the numerical solution to~(\ref{eomsphere}). The blue, red and green curves correspond to the cases $d=2,3,4$ respectively.}
\end{center}
\end{figure}

Following~\cite{andrei}, let us take the ansatz $\xi(\rho)=\lambda e^{-A\rho^{B}}$, where $A, B$ and $\lambda$ are constants. Note that in large $R$ limit, most of the hypersurface lies in the near horizon region, hence the metric~(\ref{confads2}) provides an approximate description. The value of $\lambda$ may be fixed as follows: We impose the condition that the crossover from the near-horizon region to the full metric to happen at $\rho\sim R$, which leads to $\xi(R)\sim 1$, hence $\lambda=e^{AR^{B}}$.
After substituting the ansatz for $\xi(\rho)$ to ~(\ref{eomsphere}), the equation of motion becomes
\begin{equation}
\eta\rho^{4}+\eta A^{2}B^{2}\rho^{2+2B}-A^{3}B^{3}(d-1)\rho^{3B}-AB(B+d-2)\rho^{2+B}=0.
\end{equation}
The values of $A$ and $B$ can be determined by extracting the leading order (large $R$) behavior,
\begin{equation}
B=2,~~~A=\frac{\eta}{2(d-1)},\Rightarrow \xi(\rho)=\lambda e^{-\frac{\eta}{2(d-1)}\rho^{2}}.
\end{equation}
The behavior of $\xi(\rho)$ in different dimension $d$ is plotted in Figure 3.1. It should be emphasized that there is no trivial solution $\xi^{\prime}=0$ in this case, hence the phase transition seen in the strip case cannot be observed here.

\begin{figure}
\label{nufit}
\begin{center}
\vspace{-1cm}
\hspace{-0.5cm}
\includegraphics[angle=0,width=0.45\textwidth]{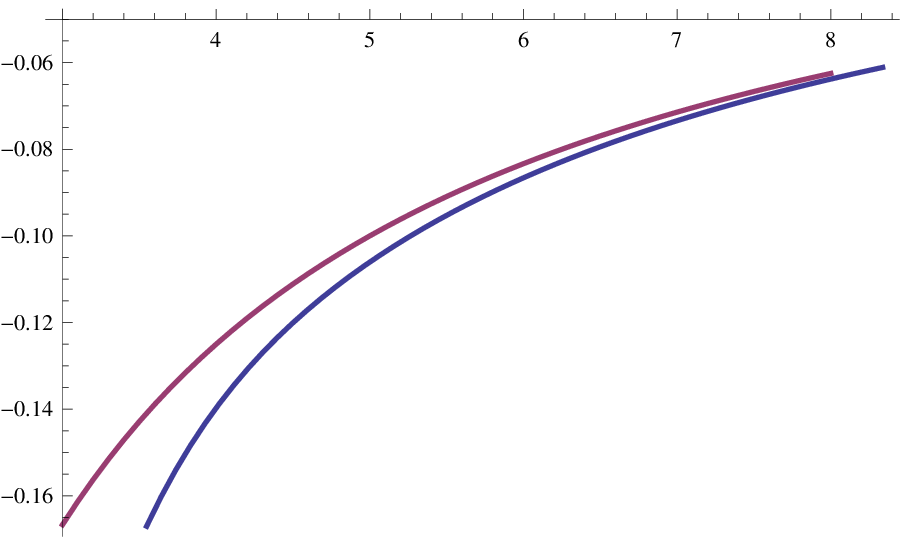}\put(-218,2){$A_{1}$}\put(-10,125){$\rho$}
\hspace{0.3cm}
\includegraphics[angle=0,width=0.45\textwidth]{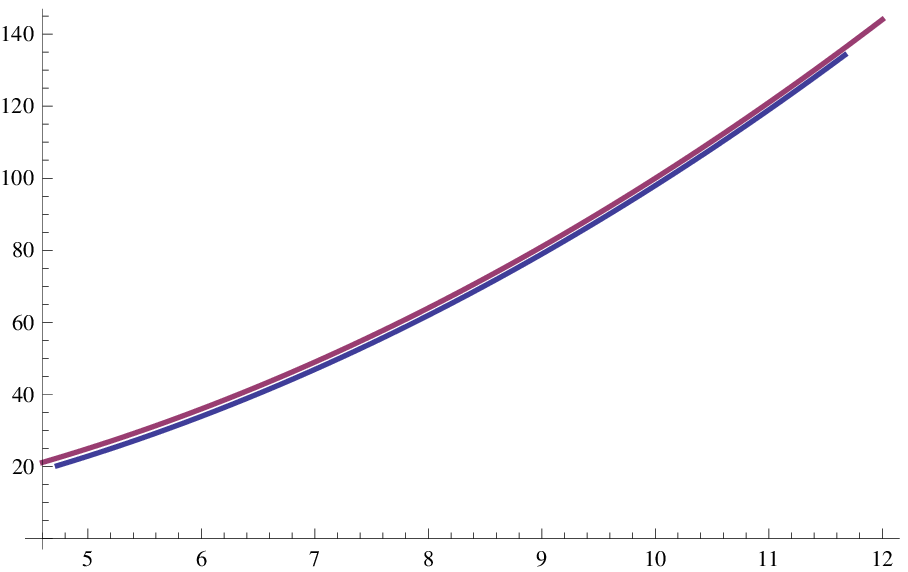}\put(-190,120){$A(\gamma)$}\put(-10,-7){$\rho$}
\vspace{-0cm} \\
\caption{\small Fitting~(\ref{leading}) with numerical results from~(\ref{Agm}) at $\eta=1$. The plot on the left is for the $d=2$ case and $A_{1}=A(\gamma)-R$. The plot on the right is for $d=3$. The blue curves denote the numerical results and the purple curves denote the leading order results in~(\ref{leading}).}
\end{center}
\end{figure}

The holographic entanglement entropy is given by
\begin{eqnarray}
S&\propto&\int d\rho\frac{\rho^{d-1}}{\xi^{\eta}}\sqrt{1+\frac{\xi^{\prime2}}{\xi^{2}}}\nonumber\\
&\simeq&\int d\rho\rho^{d}e^{A\eta\rho^{2}}.
\end{eqnarray}
It can be verified that for all dimensions $d$, the leading order term is given by $R^{d-1}$, which means that the area law always holds.
Note this is an IR behavior while the usual 'area law' refers to the UV behavior. In particular, we have the following results for $d=2,3$,
\begin{eqnarray}
\label{leading}
& &d=2,~~S\sim A(\gamma)=R-\frac{1}{R\eta^{4}},\nonumber\\
& &d=3,~~S\sim A(\gamma)={R}^{2}.
\end{eqnarray}
In Figure~\ref{nufit} we fit our leading order results~(\ref{leading}) with straightforward numerical integrations. Note that for the $d=2$ case we fit $A_{1}\equiv A(\gamma)-R=-1/R$ with the result given by taking $A(\gamma)$ to be the numerical integration in~(\ref{Agm}). the $d=3$ case we fit $A(\gamma)$ in ~(\ref{leading}) with the numerical integration in~(\ref{Agm}). It can be seen that the analytic results match the numerical results very well.

Here are some remarks on the result of $d=2$. Consider a disk with a smooth boundary of length $L$ in $(2+1)$ dimensions, the entanglement entropy is given by
\begin{equation}
S_{A}=\alpha L-\log\mathcal{D}+\cdots.
\end{equation}
The parameter $\alpha$ in the first term is cutoff dependent and the second term gives the topological entanglement entropy, where $\mathcal{D}$ denotes the total quantum dimension. The topological entanglement entropy is a constant term and it provides a measure of topological order. Holographic calculations of topological entanglement entropy in confining backgrounds were performed in~\cite{Pakman:2008ui}, where the authors observed that the topological entanglement entropy vanishes in the large $L$ limit. Coming back to our results, it can be seen that there is no topological entanglement entropy in the $d=2$ case, which might suggest that there is no long-range topological order in the ground state of semi-local quantum liquids.
\section{The AdS completion}
In this section we construct solutions which are asymptotically AdS and possess semi-locality in the IR. As argued in~\cite{Kulaxizi:2012gy}, when the full geometry is considered, for sufficiently large boundary separation length $l$ the hypersurface in the bulk should probe the IR geometry. However, there exists a critical value $l_{\rm crit}$ which corresponds to the maximal separation length for which there still is a connected hypersurface. For $l>l_{\rm crit}$ only the disconnected solution exists. The transition between the connected and disconnected hypersurfaces is second order, since the former asymptotically approaches the latter as $l\rightarrow l_{\rm crit}$. When $l$ is sufficiently small, the hypersurface probes the UV region of the geometry, and $l$ is expected to be a smooth function of the turning point $z_{\ast}$. The background studied in~\cite{Kulaxizi:2012gy} involves a two-charge dilatonic black hole, which is an exact solution of type IIB supergravity truncated on $S^{5}$ with only two of the three $U(1)$ charges being equal and nonzero. The black hole is asymptotically $AdS_{5}$ and possesses semi-locality with $\eta=1$ in the IR. Therefore in order to show that the picture works at a more general level, we should first perform the AdS completion.

Following~\cite{Ogawa:2011bz}, let us begin with the Einstein-Maxwell-dilaton action
\begin{equation}
S=\int d^{d+2}x\sqrt{-g}\left[R-\frac{1}{2}(\nabla\Phi)^{2}-V(\Phi)-\frac{1}{4}Z(\Phi)F_{\mu\nu}F^{\mu\nu}\right].
\end{equation}
The corresponding equations of motion are given by
\begin{equation}
\partial_{\mu}(\sqrt{-g}Z(\Phi)F^{\mu\nu})=0,
\end{equation}
\begin{equation}
\partial_{\mu}(\sqrt{-g}\partial^{\mu}\Phi)=\frac{1}{4}\sqrt{-g}\frac{\partial Z(\Phi)}{\partial\Phi}
F_{\rho\sigma}F^{\rho\sigma}+\sqrt{-g}\frac{\partial V}{\partial\Phi},
\end{equation}
\begin{eqnarray}
& &R_{\mu\nu}-\frac{1}{2}Rg_{\mu\nu}+\frac{1}{2}g_{\mu\nu}V(\Phi)-\frac{1}{2}\nabla_{\mu}\Phi\nabla_{\nu}\Phi
+\frac{1}{4}g_{\mu\nu}(\nabla\Phi)^{2}\nonumber\\
& &-\frac{1}{2}Z(\Phi)F_{\mu\lambda}F_{\nu}^{\lambda}+
\frac{1}{8}Z(\Phi)g_{\mu\nu}F_{\rho\sigma}F^{\rho\sigma}=0.
\end{eqnarray}

The ansatz for the solution is as follows:
\begin{equation}
\label{adsansatz}
ds^{2}_{d+2}=\frac{L^{2}}{z^{2}}\left[-f(z)dt^{2}+g(z)dz^{2}+\sum\limits^{d}_{i=1}dx_{i}^{2}\right],~~A_{t}=A_{t}(z).
\end{equation}
The Einstein tensor can be obtained by making use of~(\ref{adsansatz}),
\begin{eqnarray}
& &G_{tt}=-\frac{df(z)((d+1)g(z)+zg^{\prime}(z))}{2z^{2}g(z)^{2}},\nonumber\\
& &G_{zz}=\frac{d((d+1)f(z)-zf^{\prime}(z))}{2z^{2}f(z)},\nonumber\\
& &G_{ii}=-\frac{1}{4z^{2}f(z)^{2}g(z)^{2}}[z^{2}g(z)f^{\prime2}(z)-2df(z)^{2}\nonumber\\
& &((d+1)g(z)+zg^{\prime}(z))+zf(z)(zf^{\prime}(z)g^{\prime}(z)+2g(z)(df^{\prime}(z)-zf^{\prime}(z)))].
\end{eqnarray}
An appropriate energy condition should be imposed in order to have a physically sensible solution, so here we consider the null energy condition (NEC), $T_{\mu\nu}N^{\mu}N^{\nu}\geq0$, where $N^{\mu}$ denotes any null vector and $T_{\mu\nu}=G_{\mu\nu}$.  We can take the following components of the null vector,
\begin{equation}
N^{t}=\frac{1}{\sqrt{f(z)}},~~N^{z}=\frac{\cos\theta}{\sqrt{g(z)}},~~N^{x}=\sin\theta,
\end{equation}
where $\theta$ is an arbitrary constant. Then it can be seen that
\begin{eqnarray}
T_{\mu\nu}N^{\mu}N^{\nu}&=&-\frac{\sin^{2}\theta}{4zf(z)^{2}g(z)^{2}}[zg(z)f^{\prime2}(z)\nonumber\\
& &+f(z)\left(zf^{\prime}(z)g^{\prime}(z)+g(z)\left(2df^{\prime}(z)-2zf^{\prime\prime}(z)\right)\right)]\nonumber\\
& &-\cos^{2}\theta\frac{d(g(z)f^{\prime}(z)+g^{\prime}(z)f(z))}{2zf(z)g(z)^{2}}.
\end{eqnarray}
The NEC is satisfied if and only if
\begin{equation}
g(z)f^{\prime}(z)+g^{\prime}(z)f(z)\leq0,
\end{equation}
\begin{equation}
zg(z)f^{\prime2}(z)+f(z)\left(zf^{\prime}(z)g^{\prime}(z)
+g(z)\left(2df^{\prime}(z)-2zf^{\prime\prime}(z)\right)\right)\leq0,
\end{equation}

Note that we are looking for asymptotically $AdS_{d+2}$ solutions where the AdS boundary is located at $z=0$, so
$f(0)=g(0)=1$. Moreover, we introduce a scale $z_{F}$ such that $z\gg z_{F}$ corresponds to the IR limit
and $z\ll z_{F}$ corresponds to the UV limit. The solution for the $U(1)$ gauge field is easily obtained by substituting the background metric into the equation of motion,
\begin{equation}
A_{t}^{\prime}(z)=\frac{A}{Z(\Phi)}\sqrt{f(z)g(z)}z^{d-2},
\end{equation}
where $A$ is the integration constant. The solutions for $\Phi, V(\Phi), Z(\Phi)$ are as follows,
\begin{eqnarray}
V(\Phi)&=&\frac{1}{4L^{2}f(z)^{2}g(z)^{2}}[z^{2}g(z)f^{\prime2}(z)+2df(z)^{2}(2(d+1)g(z)+zg^{\prime}(z))\nonumber\\
& &+zf(z)(zf^{\prime}(z)g^{\prime}(z)+2g(z)(2df^{\prime}(z)-zf^{\prime\prime}(z)))],
\end{eqnarray}
\begin{equation}
\Phi^{\prime2}=-\frac{d(g(z)f^{\prime}(z)+f(z)g^{\prime}(z))}{zf(z)g(z)},
\end{equation}
\begin{eqnarray}
\frac{1}{Z(\Phi)}&=&-\frac{L^{2}}{2A^{2}f(z)^{2}g(z)^{2}z^{2d-1}}[zg(z)f^{\prime2}(z)+\nonumber\\
& &f(z)(zf^{\prime}(z)g^{\prime}(z)+2g(z)(df^{\prime}(z)-zf^{\prime\prime}(z)))],
\end{eqnarray}
Note that if we impose physical constraints $\Phi^{\prime2}\geq0,~~Z(\Phi)\geq0$, the above equations lead to exactly the same expressions as those derived from NEC.

For completeness we first consider the IR solution,
\begin{equation}
\label{sec4IR}
f(z)=kz^{-p},~~g(z)=\frac{z_{F}^{2}}{z^{2}},
\end{equation}
where $p\equiv2d/\eta$ and $k$ is a positive constant.
The solution for $V(\Phi), \Phi$ and $Z(\Phi)$ are given by
\begin{equation}
V(\Phi)=-\frac{(p+2d)^{2}z^{2}}{4L^{2}z_{F}^{2}},~~\Phi^{\prime2}=\frac{d(p+2)}{z^{2}},~~~Z(\Phi)=
\frac{2A^{2}z_{F}^{2}z^{2d-2}}{L^{2}p(p+2d)}.
\end{equation}
We write $V(\Phi)$ and $Z(\Phi)$ in terms of $\Phi$,
\begin{eqnarray}
\Phi&=&\sqrt{d(p+2)}\log z,\nonumber\\
V(\Phi)&=&-\frac{(p+2d)^{2}}{4L^{2}z_{F}^{2}}e^{\frac{2\Phi}{\sqrt{d(p+2)}}},\nonumber\\
Z(\Phi)&=&\frac{2A^{2}z_{F}^{2}}{p(p+2d)L^{2}}e^{\frac{2(d-1)\Phi}{\sqrt{d(p+2)}}}.
\end{eqnarray}
The black hole solution reads
\begin{equation}
g(z)=\frac{z_{F}^{2}}{z^{2}h(z)},~~f(z)=\frac{k}{z^{p}}h(z),~~h(z)=1-(\frac{z}{z_{H}})^{d+p/2},
\end{equation}
while the other field configurations remain the same.

\begin{figure}
\label{VZvsz}
\begin{center}
\vspace{-1cm}
\hspace{-0.5cm}
\includegraphics[angle=0,width=0.45\textwidth]{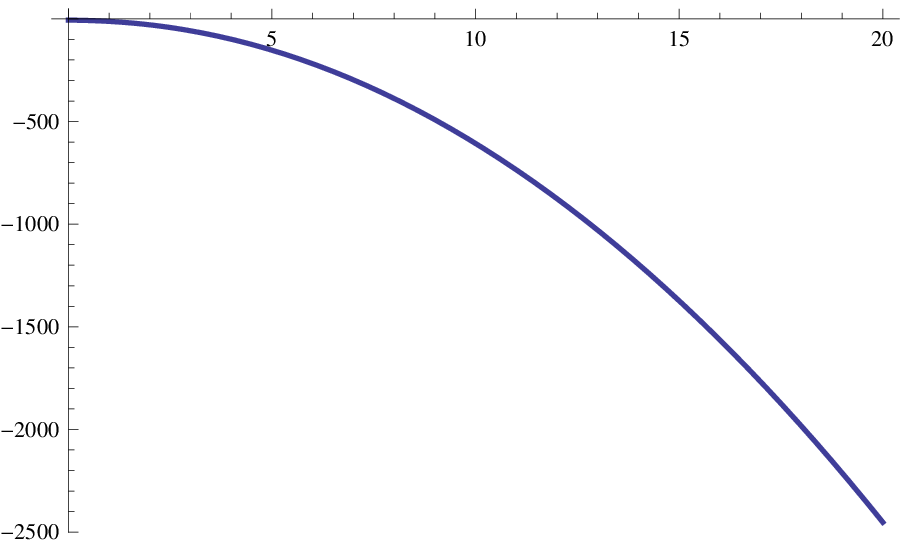}\put(-184,2){$V$}\put(-10,125){$z$}
\hspace{0.3cm}
\includegraphics[angle=0,width=0.45\textwidth]{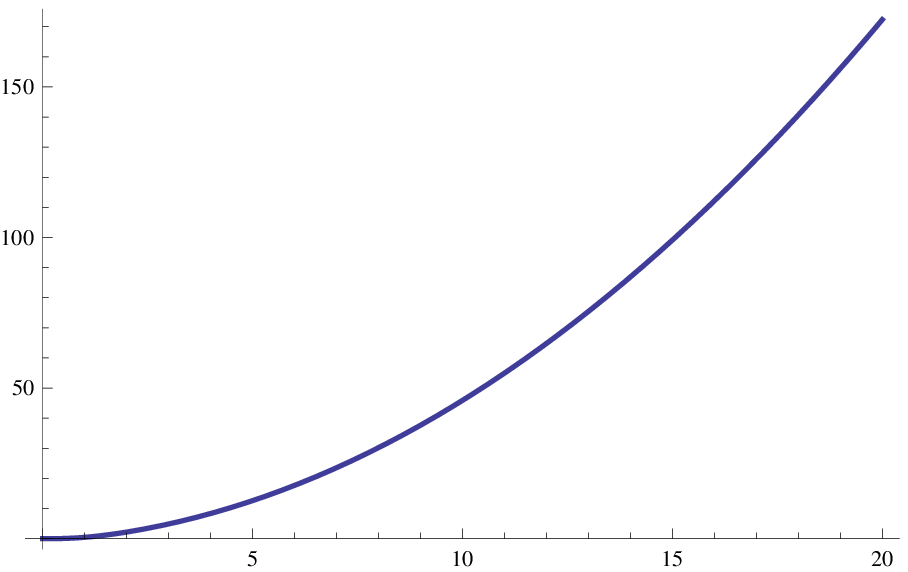}\put(-190,115){$1/Z$}\put(-10,-7){$z$}
\vspace{-0cm} \\
\caption{\small The plots for $V(\Phi)$ and $Z(\Phi)$ with $d=2$. $V(\Phi)$ reproduces the value of the cosmological constant at the leading order in the UV $z\rightarrow0$ and blows up in the IR $z\rightarrow0$, which mimics the behavior of the exponential scalar potential. $Z(\Phi)$ is strictly positive in the whole geometry. We set $p=k=L=z_{F}=1$.}
\end{center}
\end{figure}

Next we will embed the zero-temperature IR solution~(\ref{sec4IR}) into AdS spacetime. We may take the following ansatz for $f(z)$ and $g(z)$,
\begin{equation}
f(z)=\frac{k}{k+z^{p}},~~g(z)=\frac{z_{F}^{2}}{z^{2}+z_{F}^{2}}.
\end{equation}
It can be seen that $f(0)=g(0)=1$ and $f(z), g(z)$ reduce to the IR solution~(\ref{sec4IR}) when $z\rightarrow\infty$. Then the solutions for $\Phi$ and $Z(\Phi)$ are given by
\begin{equation}
\Phi^{\prime2}=\frac{d[2kz^{2}+(p+2)z^{p+2}+pz_{F}^{2}z^{p}]}{z^{2}(k+z^{p})(z^{2}+z_{F}^{2})},
\end{equation}
\begin{equation}
\frac{1}{Z(\Phi)}=\frac{L^{2}pz^{p-2d}}{2A^{2}(k+z^{p})^{2}z_{F}^{2}}[(z^{2}+z_{F}^{2})(2d(k+z^{p})-
2kp+pz^{p})+2(k+z^{p})z_{F}^{2}].
\end{equation}
Note that $\Phi^{\prime2}$ is always $\geq0$ while it is not the case for $Z(\Phi)$. However, we may impose a sufficient but not necessary condition $2kd-2kp>0$ so that $1/Z(\Phi)>0$, which leads to $p<d$. The UV behavior of these fields may be obtained by taking $z\rightarrow0$,
\begin{equation}
\Phi\simeq\frac{\sqrt{2d}}{z_{F}}z,~~Z(\Phi)\simeq\frac{A^{2}k}{L^{2}p(d+1-p)}
\frac{z_{F}^{2d-p}}{(2d)^{d-p/2}}\Phi^{2d-p}.
\end{equation}
On the other hand, the scalar potential is given by
\begin{eqnarray}
V(\Phi)&=&-\frac{1}{4L^{2}(k+z^{p})^{2}z_{F}^{2}}[4k^{2}d(d+1)z_{F}^{2}+4d^{2}k^{2}z^{2}\nonumber\\
& &+2k(4d^{2}-p(p-1)+2d(p+2))z_{F}^{2}z^{p}+2k(4d^{2}+2dp-p^{2})z^{p+2}\nonumber\\
& &+(p+2d)(p+2d+2)z_{F}^{2}z^{2p}+(p+2d)^{2}z^{2p+2}],
\end{eqnarray}
If we take the UV limit $z\rightarrow0$, it becomes
\begin{equation}
V(\Phi)=-\frac{d(d+1)}{L^{2}}-\frac{d}{2L^{2}}\Phi^{2}.
\end{equation}
The first term is just the cosmological constant and the second term gives the mass square $m^{2}=-d$.  Note that the BF bound in $AdS_{d+2}$ is $m^{2}\geq-(d+1)^{2}/4$,
so the BF bound is satisfied in our case. The behavior of $V(\Phi)$ and $Z(\Phi)$ with $d=2$ is plotted in Figure~\ref{VZvsz}.

\section{HEE for the full solution}
After constructing the full solution which is asymptotically AdS in the UV and possesses semi-locality in the IR, we consider the holographic entanglement entropy for this geometry. We find that for a strip entangling region, the behavior of the entanglement entropy agrees with the picture proposed in~\cite{Kulaxizi:2012gy}, i.e. the boundary separation length $l$ is a smooth function of the turning point and it approaches $l_{\rm crit}$ as $z_{\ast}$ is large enough. As a result, the connected surface dominates when $l$ is sufficiently small and the disconnected surface dominates for $l$ which is large enough. We also consider the cases in which the entangling region is a sphere and an annulus.
\subsection{The strip}
\label{sec:strip}
\begin{figure}
\label{lvszast}
\begin{center}
\vspace{-1cm}
\hspace{-0.5cm}
\includegraphics[angle=0,width=0.45\textwidth]{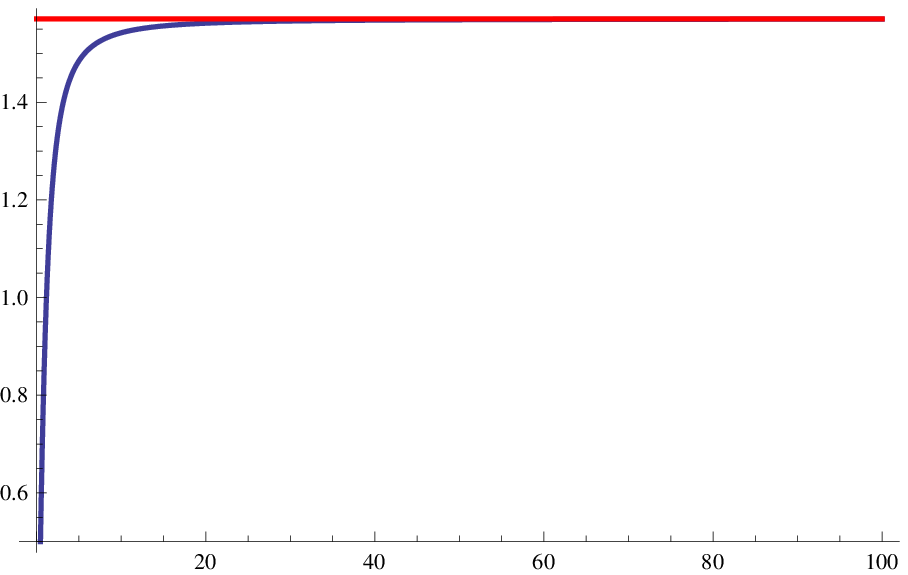}\put(-200,130){$l$}\put(-10,-7){$z_{\ast}$}
\hspace{-0cm}
\includegraphics[angle=0,width=0.45\textwidth]{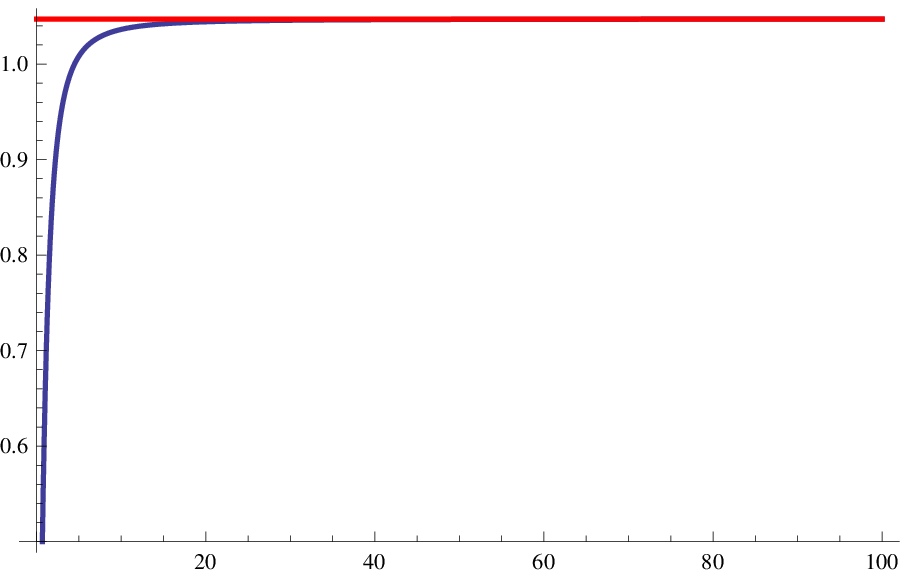}\put(-200,130){$l$}\put(-10,-7){$z_{\ast}$}
\vspace{-0cm} \\
\caption{\small The boundary separation length in the full solution (blue curve) and the IR solution (red curve). The plot on the left hand side is for $d=2$ and the one on the right hand side is for $d=3$. For both cases $l$ and $l_{\rm crit}$ have significant differences when $z_{\ast}$ is sufficiently small, which means that the minimal surface just probes the geometry near the UV. As $z_{\ast}$ increases, the minimal surface goes deeper into the IR and $l$ approaches $l_{\rm crit}$.}
\end{center}
\end{figure}

Let us consider strip case,
\begin{equation}
x_{1}\equiv x\in\left[-\frac{l}{2},\frac{l}{2}\right],~~x_{i}\in[0,L_{x}], ~i=2,\cdots,d,
\end{equation}
where $l\ll L_{x}$. The induced metric can be read off from the solution~(\ref{adsansatz})
\begin{equation}
ds^{2}_{\rm ind}=\frac{L^{2}}{z^{2}}\left((g(z)+x^{\prime2})dz^{2}+\sum\limits^{d}_{i=1}dx_{i}^{2}\right),
\end{equation}
where we have parameterized $x=x(z)$. The minimal surface area is given by
\begin{eqnarray}
A(\gamma)&=&2\int\frac{L^{d}}{z^{d}}\sqrt{g(z)+x^{\prime2}}\nonumber\\
&=&2L^{d}L_{x}^{d-1}\int\frac{dz}{z^{d}}\sqrt{g(z)+x^{\prime2}}.
\end{eqnarray}
We get the conserved quantity
\begin{equation}
C\equiv\frac{x^{\prime}}{z^{d}\sqrt{g(z)+x^{\prime2}}},
\end{equation}
which leads to
\begin{equation}
x^{\prime}=\frac{\sqrt{g(z)}(\frac{z}{z_{\ast}})^{d}}{\sqrt{1-(\frac{z}{z_{\ast}})^{2d}}},
\end{equation}
where $z_{\ast}$ denotes the turning point.
The boundary separation length is given by
\begin{equation}
\frac{l}{2}=\int^{z_{\ast}}_{0}dz\frac{\sqrt{g(z)}(\frac{z}{z_{\ast}})^{d}}{\sqrt{1-(\frac{z}{z_{\ast}})^{2d}}}.
\end{equation}
Recall that in the IR, $g(z)=z_{F}^{2}/z^{2}$, hence
\begin{equation}
l=l_{\rm crit}=\frac{\pi z_{F}}{d},
\label{eq:lcrit}
\end{equation}
which is constant. If we consider the full solution $g(z)=z_{F}^{2}/(z^{2}+z_{F}^{2})$, the boundary separation length reads
\begin{equation}
l=2\int^{z_{\ast}}_{0}dz\frac{z_{F}}{\sqrt{z^{2}+z_{F}^{2}}\sqrt{(\frac{z_{\ast}}{z})^{2d}-1}}.
\end{equation}
The behavior of $l$ vs $z_{\ast}$ is plotted in~\ref{lvszast} for $d=2,3$ with $z_{F}=1$.

It can be seen that $l$ is a smooth function of $z_{\ast}$. When $z_{\ast}$ is small, significant differences between $l$ and $l_{\rm crit}$ can be observed. However, when $z_{\ast}$ is sufficiently large, $l$ approaches $l_{\rm crit}$.

Recall that the holographic entanglement entropy is determined by
\begin{equation}
S=\frac{A(\gamma)}{4G_{N}^{(d+2)}},
\end{equation}
where $A(\gamma)$ denotes the minimal surface area. We will plot the finite part
\begin{eqnarray}
A_{\rm finite}&=&\frac{1}{2L^{d}L_{x}^{d-1}}(A(\gamma)-A_{\rm div})\nonumber\\
&=&\int^{z_{\ast}}_{0}\frac{dz}{z^{d}}\frac{\sqrt{g(z)}}{1-(\frac{z}{z_{\ast}})^{2d}}
-\frac{1}{(d-1)a^{d-1}},
\end{eqnarray}
while taking the limit $a\rightarrow0$. Note that the divergent term is the standard result obtained in~\cite{Ryu:2006ef}.
\begin{figure}
\begin{center}
\vspace{1cm}
\hspace{-0cm}
\includegraphics[angle=0,width=0.45\textwidth]{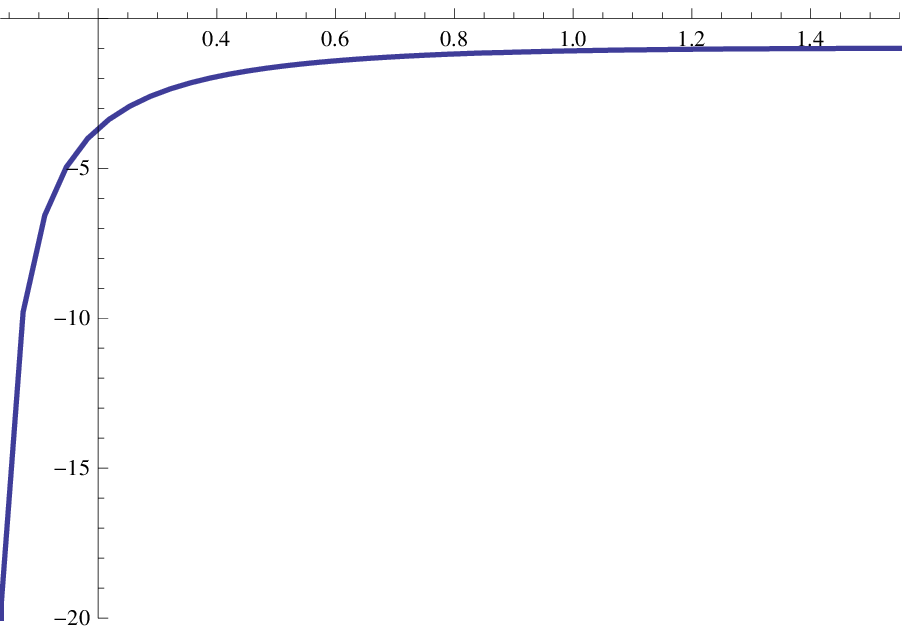}\put(-178,2){$A_{\rm finite}$}\put(-10,143){$l$}
\hspace{0.3cm}
\includegraphics[angle=0,width=0.45\textwidth]{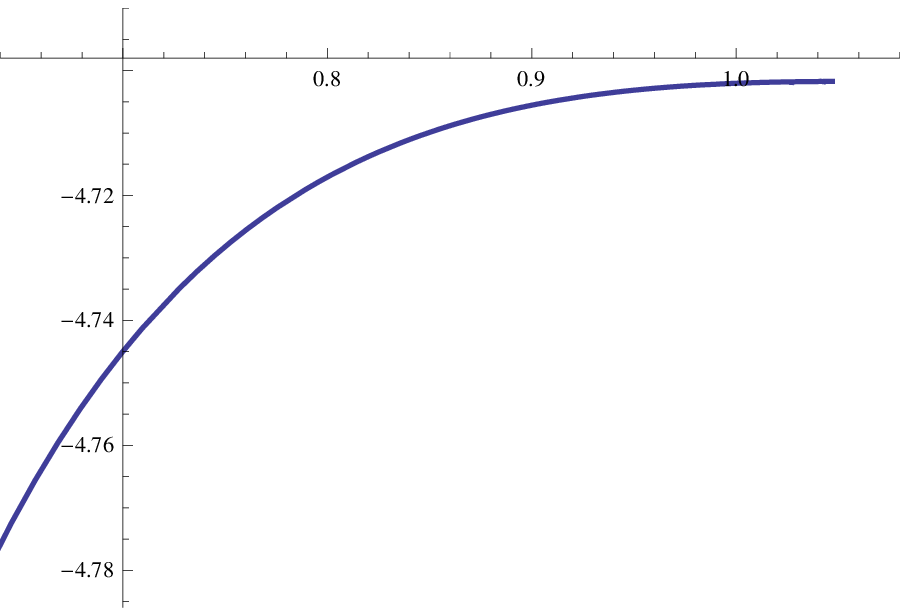}\put(-174,0){$A_{\rm finite}$}\put(-10,128){$l$}
\caption{\small The finite part of the entanglement entropy. The plot on the left is for $d=2$ and the one on the right is for $d=3$. As $l\rightarrow l_{\rm crit}$ the entanglement entropy tends to be constant.}
\label{Afinite}
\end{center}
\end{figure}
On the other hand, the disconnected surface is given by $x^{\prime}=0$, so the minimal surface area reads
\begin{equation}
A_{\rm dis}(\gamma)=L^{d}L_{x}^{d-1}\int\frac{dz}{z^{d}\sqrt{g(z)}}.
\end{equation}
The behavior of $\Delta A=A_{\rm finite}-A_{\rm disfinite}$ for $d=2,3$ is plotted in Figure ~\ref{deltaa}, where we have subtracted the
divergent term for $A_{\rm dis}(\gamma)$.
\begin{figure}
\label{deltaa}
\begin{center}
\vspace{-1cm}
\hspace{-0.5cm}
\includegraphics[angle=0,width=0.45\textwidth]{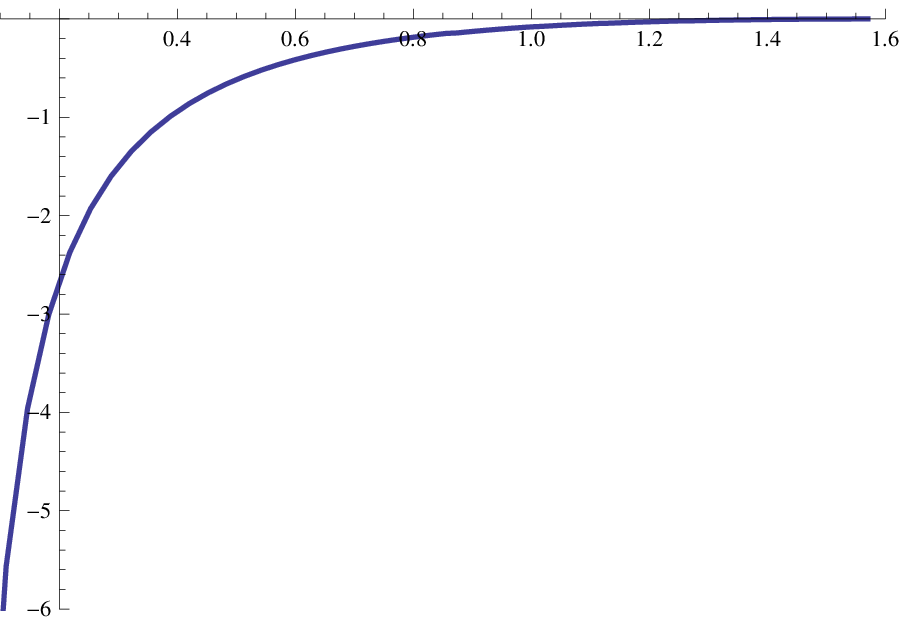}\put(-188,2){$\Delta A$}\put(-10,140){$l$}
\hspace{0.3 cm}
\includegraphics[angle=0,width=0.45\textwidth]{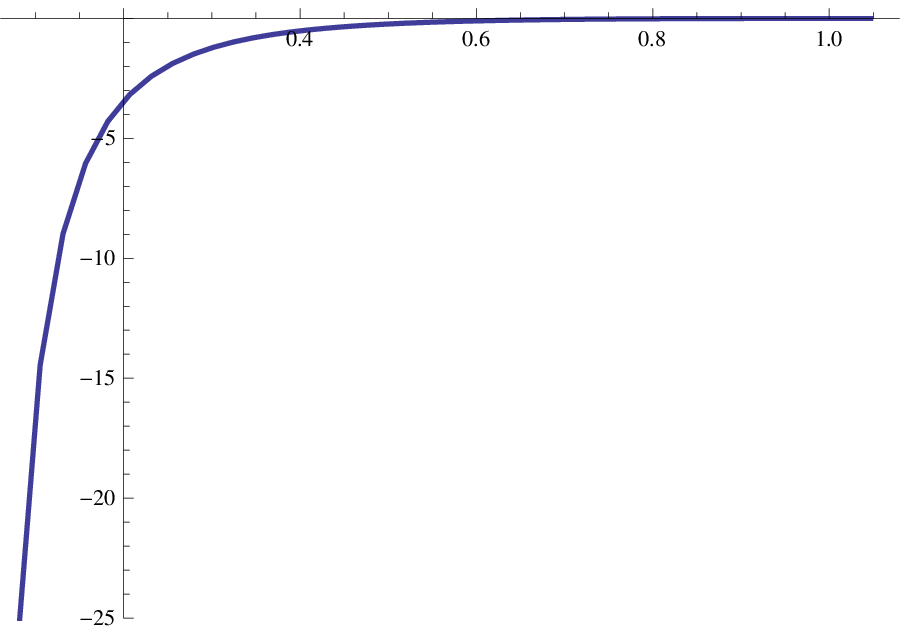}\put(-174,2){$\Delta A$}\put(-10,142){$l$}
\caption{\small The differences between the HEE of the connected minimal surface and the disconnected one. The plot on the left is for $d=2$ and the one on the right is for $d=3$. When $l$ is sufficiently small, the connected minimal surface dominates. As $l\rightarrow l_{\rm crit}$ the disconnected one dominates.}
\end{center}
\end{figure}
It can be seen that when $l<l_{\rm crit}$, the connected surface dominates,
as $l\rightarrow l_{\rm crit}$, the difference tends to zero, which signifies that the disconnected surface will dominate. This behavior agrees with the picture proposed in~\cite{Kulaxizi:2012gy}.
\subsection{The sphere}
\begin{figure}
\label{rhovsz}
\begin{center}
\vspace{-1cm}
\hspace{-0.5cm}
\includegraphics[angle=0,width=0.45\textwidth]{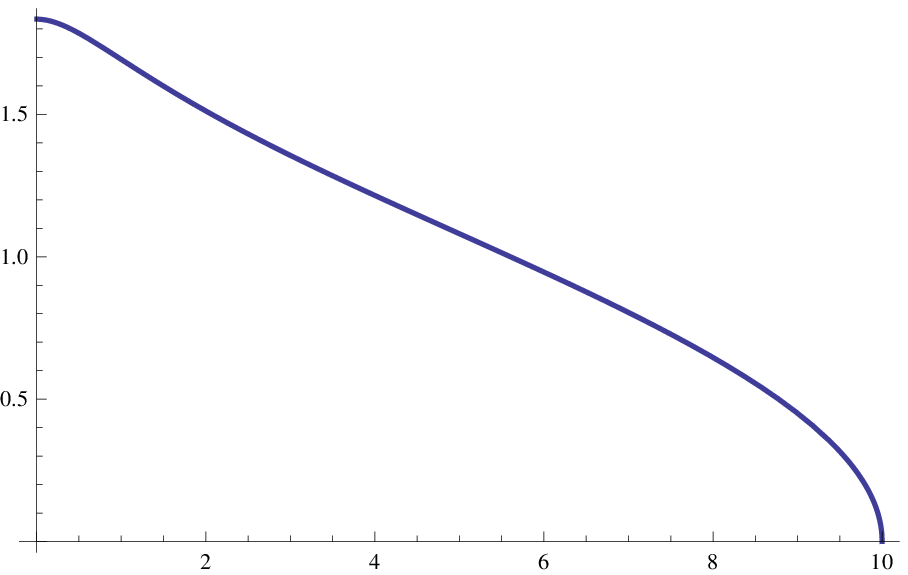}\put(-200,130){$\rho$}\put(-10,-7){$z$}
\hspace{-0cm}
\includegraphics[angle=0,width=0.45\textwidth]{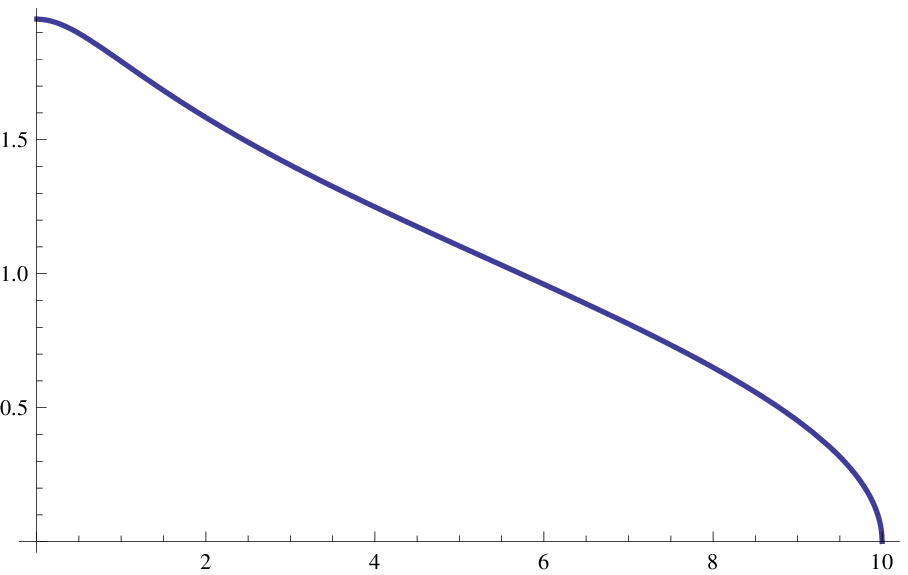}\put(-200,130){$\rho$}\put(-10,-7){$z$}
\vspace{0.5cm}\\
\caption{\small The profile of $\rho$, left $d=2$ right $d=3$.}
\label{rho}
\end{center}
\end{figure}

Next we consider the case of a spherical entangling region with $\rho\in[0,R]$. The induced metric is given by
\begin{equation}
ds^{2}_{\rm ind}=\frac{L^{2}}{z^{2}}[(g(z)+\rho^{\prime2})dz^{2}+\rho^{2}d\Omega^{2}_{d-1}].
\end{equation}
The minimal surface area is
\begin{equation}
A_{\rm sphere}=L^{d}{\rm Vol}(\Omega_{d-1})\int\frac{dz}{z^{d}}\rho^{d-1}\sqrt{g(z)+\rho^{\prime2}},
\end{equation}
from which we can derive the equation of motion for $\rho(z)$,
\begin{equation}
\partial_{z}\left(\frac{\rho^{d-1}\rho^{\prime}}{z^{d}\sqrt{g(z)+\rho^{\prime2}}}\right)
=\frac{(d-1)\rho^{d-2}}{z^{d}}\sqrt{g(z)+\rho^{\prime2}}.
\end{equation}
Note that in this case there is no conserved quantity or trivial solution $\rho^{\prime}=0$. We can solve for $\rho(z)$ numerically by fixing the boundary conditions $\rho(0)=R, \rho(z_{\ast})=0$, where $z_{\ast}$ denotes the turning point. The plots for $d=2$ and $d=3$ are shown in Figure~\ref{rhovsz}.
We plot the finite part of the holographic entanglement entropy
\begin{equation}
A_{\rm finite}=\frac{1}{L^{d}{\rm Vol}(\Omega_{d-1})}(A_{\rm sphere}-A_{\rm div}),
\end{equation}
where
\begin{eqnarray}
& &A_{\rm div}=\frac{R}{a},~~~d=2,\nonumber\\
& &A_{\rm div}=\frac{R^{2}}{2a^{2}}+\frac{1}{2}\log a,~~~d=3,
\end{eqnarray}
are the standard results given in~\cite{Ryu:2006ef}.
\begin{figure}
\begin{center}
\vspace{-1cm}
\hspace{-0.5cm}
\includegraphics[angle=0,width=0.45\textwidth]{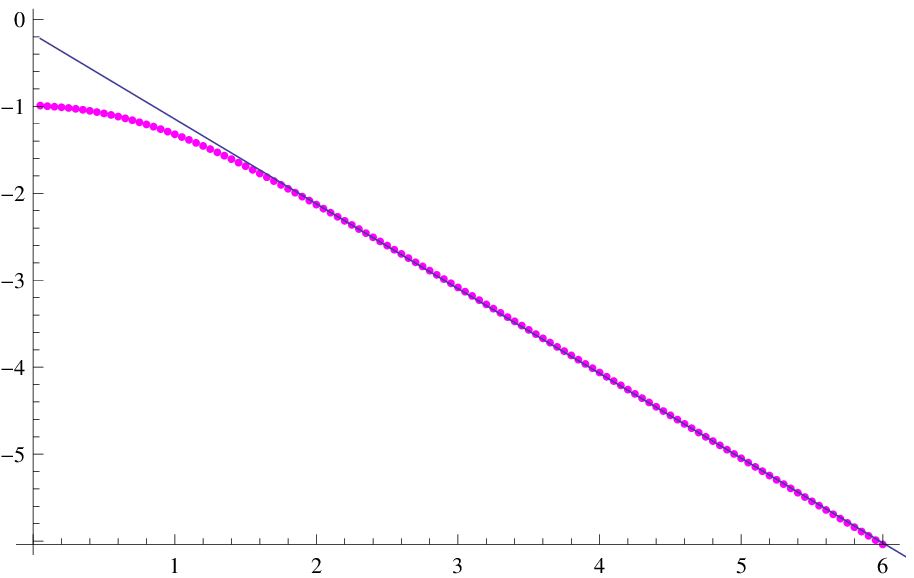}\put(-200,130){$A_{\rm finite}$}\put(-10,-9){$R$}
\hspace{-0cm}
\includegraphics[angle=0,width=0.45\textwidth]{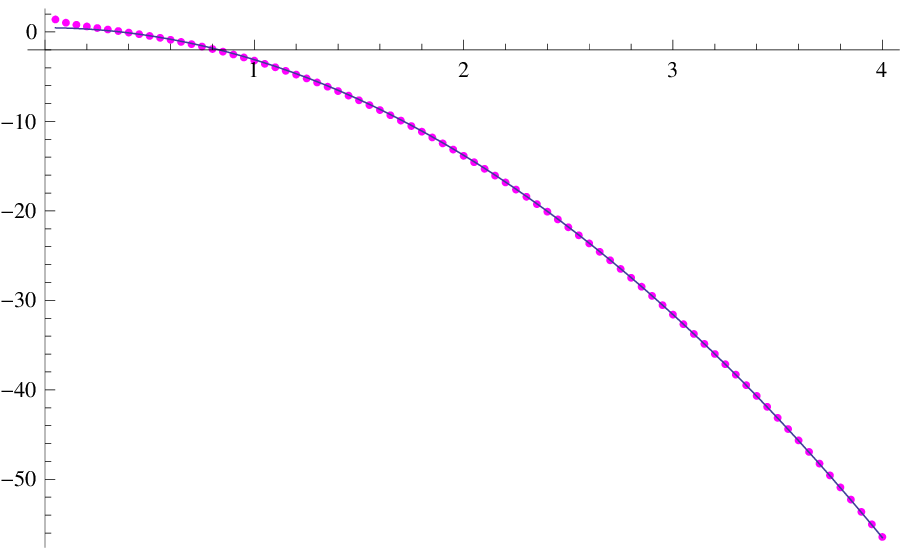}\put(-200,130){$A_{\rm finite}$}\put(-10,120){$R$}
\vspace{0.5cm}\\
\caption{\small The finite part with spherical entangling region. The plot on the left is for $d=2$ and the one on the right is for $d=3$. The dots are data from numerical evaluations and the curves denote the fits. }
\label{Afsphere}
\end{center}
\end{figure}
We are interested in the deviation of the finite part of HEE from the area law~\cite{Kulaxizi:2012gy}, which can be analyzed by performing the numerical fits on the numerical data. The resulting behavior reads
\begin{eqnarray}
& &A_{\rm finite}=-0.171363-0.974893R\ \text{ for }\ d=2,\nonumber\\
& &A_{\rm finite}=0.469379-3.56108R^2\ \text{ for }\ d=3,
\end{eqnarray}
which may indicate that for large $R$ the finite HEE is still governed by the area law, consistent with the conclusion in~\cite{Kulaxizi:2012gy}.

\subsection{The annulus}
From our evaluation of the holographic entanglement entropy for the cases of a strip and a sphere we may conclude that there is a phase transition in the strip case, while no such transition occurs for the sphere case. This behavior has also been observed in~\cite{Kulaxizi:2012gy}, where the background is a charged dilatonic black hole in type IIB supergravity truncated on $S^{5}$, whose near horizon geometry possesses semi-locality with $\eta=1$.

As argued in~\cite{Kulaxizi:2012gy}, a third scale supplied by the anisotropy of the strip should play a role in understanding the phase transition. One way to see this is to consider deforming the sphere entangling surface continuously into an ellipsoid, which can finally result in a strip shape entangling region. The phase transition should appear suddenly during this process. However, the ellipsoid is technically quite complex, hence we focus on a simpler case, the annulus, and leave the ellipsoid to future work.

In the annulus case we expect to approximate a sphere in the limit of vanishing inner radius and the strip for both, the inner and outer radius, large in comparison to their difference. We will see that this interpolation between these two geometries will not work out entirely as expected. In the following we will calculate the holographic entanglement entropy for annulus entangling region.

In this case we parametrize $z=z(\rho)$ and obtain the induced metric
\begin{equation}
ds^{2}_{\rm ind}=\frac{L^{2}}{z^{2}}[(1+g(z)\dot{z}^{2})d\rho^{2}+\rho^{2}d\Omega^{2}_{d-1}],
\end{equation}
where dot denotes partial derivative with respect to $\rho$.
The minimal surface area is given by
\begin{equation}
A_{\rm ann}=L^{d}{\rm Vol}(\Omega_{d-1})\int d\rho\frac{\rho^{d-1}}{z^{d}}
\sqrt{1+g(z)\dot{z}^{2}},
\end{equation}
which leads to the equation of motion
\begin{equation}
\partial_{\rho}\left(\frac{\rho^{d-1}g(z)\dot{z}}{z^{d}\sqrt{1+g(z)\dot{z}^{2}}}\right)
=-\frac{d\rho^{d-1}}{z^{d+1}}\sqrt{1+g(z)\dot{z}^{2}}
+\frac{\rho^{d-1}}{2z^{d}}\frac{\dot{z}^{2}}{\sqrt{1+g(z)\dot{z}^{2}}}\partial_{z}g(z),
\label{eq:eomannulus}
\end{equation}
with boundary condition $z(\rho_{1})=z(\rho_{2})=a\to0$.
We plot the finite part of the minimal surface area
\begin{equation}
A_{\rm finite}=\frac{1}{L^{d}{\rm Vol}(\Omega_{d-1})}(A_{\rm ann}-A_{\rm div}),
\end{equation}
where the divergent terms are given by~\cite{Hirata:2006jx},
\begin{eqnarray}
& &A_{\rm div}=\frac{\rho_{1}+\rho_{2}}{a},~~~d=2,\nonumber\\
& &A_{\rm div}=\frac{\rho_{1}^{2}+\rho_{2}^{2}}{2a^{2}}-\frac{1}{2}
\log\frac{\rho_{1}\rho_{2}}{a^{2}},~~~d=3.
\end{eqnarray}

\begin{figure}
\centering
\vspace{-1cm}
\hspace{-0.5cm}
\includegraphics[angle=0,width=0.45\textwidth]{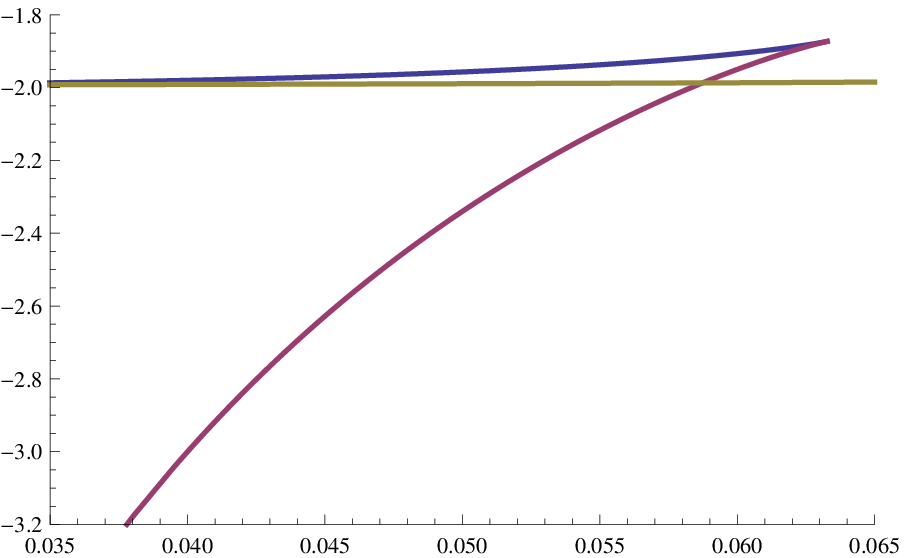}\put(-188,115){$A_{\rm finite}$}\put(-10,-9){$\Delta \rho$}
\hspace{0.3cm}
\includegraphics[angle=0,width=0.45\textwidth]{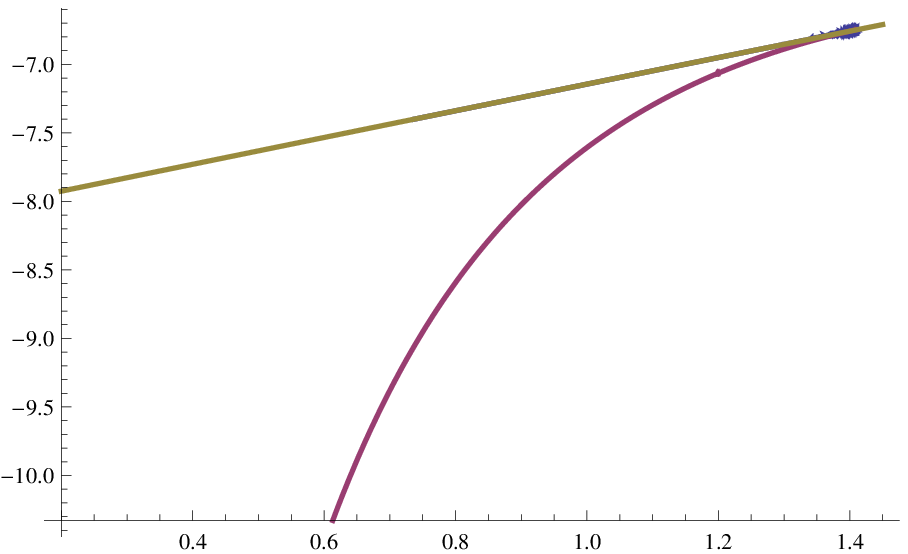}\put(-186,115){$A_{\rm finite}$}\put(-10,-9){$\Delta \rho$}
\caption{\small Finite part of the annulus entangling region for $d=2$ versus the difference of the radii $\Delta \rho = \rho_2-\rho_1$. The left plot has $\rho_2 = 0.1$ and the right one $\rho_2=4$ in terms of $z_F$. Note that for small differences $\Delta \rho \leq (\Delta \rho)_\text{max}$ we see different solutions, two connected (deformed annulus) solutions with the lower one being preferred (blue and red) and the concentric balls solution (yellow). The transition between the connected and disconnected solutions at $(\Delta \rho)_\text{crit}$ is first order for small values of $\rho_2$ as becomes obvious from the swallow tail form of the left plot. For larger values of $\rho_2$ we have a second order transition (see right plot). If $\Delta \rho>(\Delta \rho)_\text{max}$ the disconnected solution is the only solution, this behaviour is very similar to the strip case discussed in section \ref{sec:strip}. To generate these plots we set $z_F=1$ and the cutoff $a=0.001$.}
\label{Afanulus2}
\end{figure}

\begin{figure}
\centering
\hspace{-0.5cm}
\includegraphics[angle=0,width=0.45\textwidth]{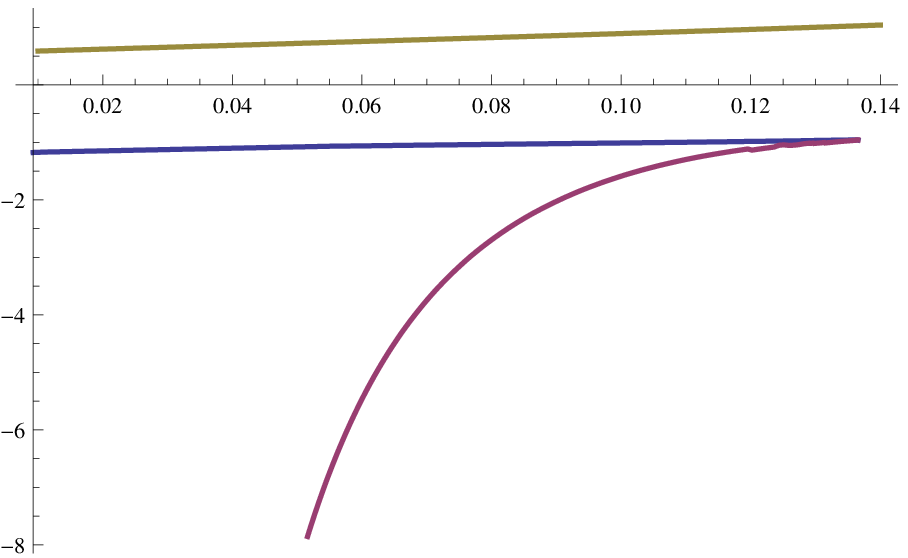}\put(-230,115){$A_{\rm finite}$}\put(-13,110){$\Delta \rho$}
\hspace{0.3cm}
\includegraphics[angle=0,width=0.45\textwidth]{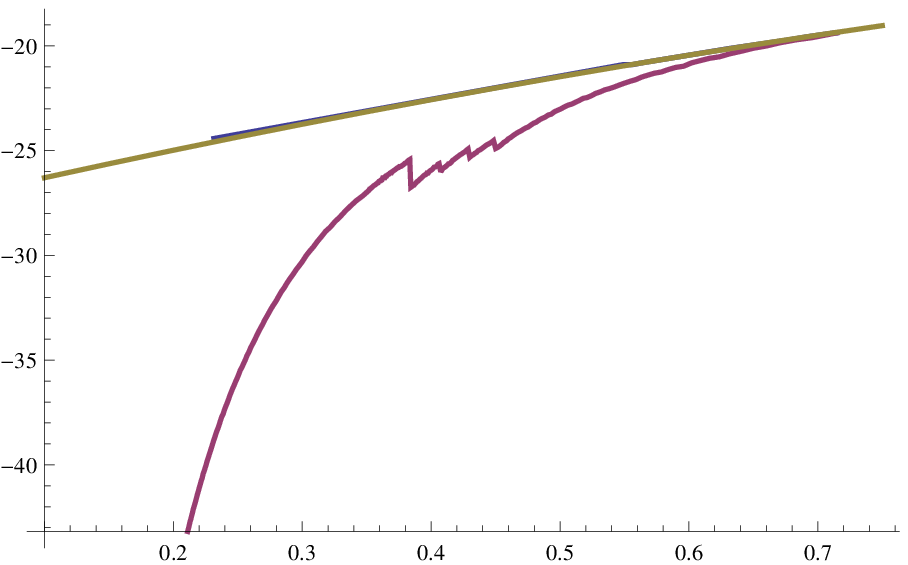}\put(-190,115){$A_{\rm finite}$}\put(-10,-7){$\Delta \rho$}
\caption{\small Finite part of the annulus entangling region for $d=3$ versus the difference of the radii $\Delta \rho = \rho_2-\rho_1$. The left plot has $\rho_2 = 0.3$ and the right one $\rho_2=2$ in terms of $z_F$. Note that for small differences $\Delta \rho \leq (\Delta \rho)_\text{crit}$ we see different solutions, two connected (deformed annulus) solutions with the lower one being preferred (blue and red) and the concentric balls solution (yellow). In contrast to the $d=2$ case we don't find a transition for small values of $\rho_2<\mathcal{O}(1)$ and a second order one for larger values. For larger values of $\Delta \rho$ the disconnected solution is the only solution, this behaviour is very similar to the strip case discussed in section \ref{sec:strip}. To generate these plots we set $z_F=1$ and the cutoff $a=0.001$. The jagged feature in the right plot is an artifact of the numerical computation and has no physical interpretation.}
\label{Afanulus3}
\end{figure}

\begin{figure}
\centering
\vspace{-1cm}
\hspace{-0.5cm}
\includegraphics[angle=0,width=0.45\textwidth]{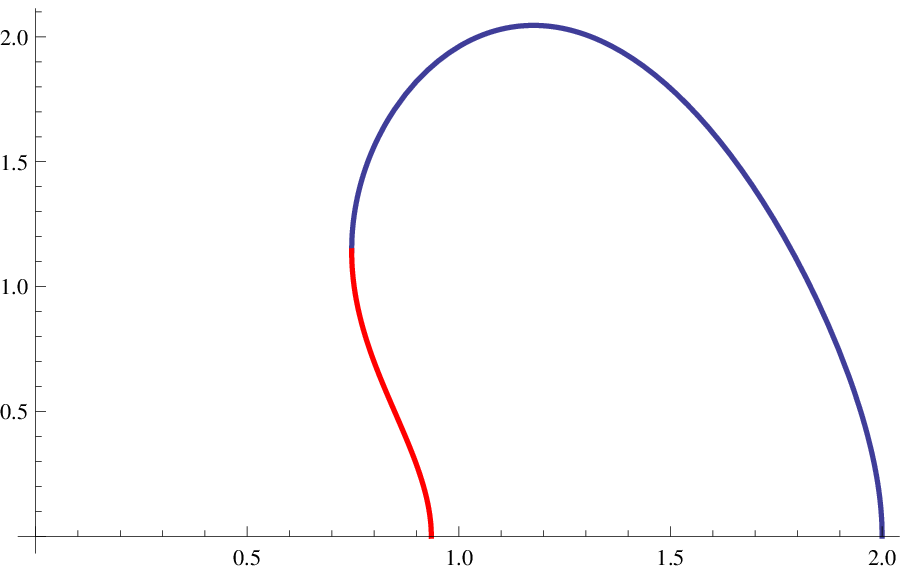}\put(-200,130){$z(\rho)$}\put(-10,-7){$\rho$}
\hspace{-0cm}
\includegraphics[angle=0,width=0.45\textwidth]{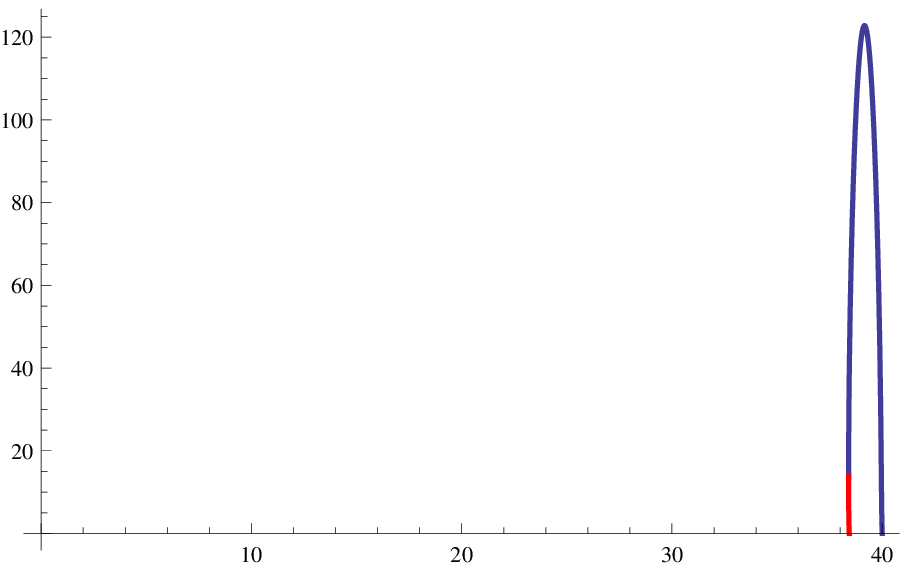}\put(-200,130){$z(\rho)$}\put(-10,-7){$\rho$}
\vspace{0.5cm}\\
\caption{\small Generic connected (annulus) solution. $z$ is the radial AdS coordinate and $\rho$ the radius of the spherical coordinates on the boundary. Both plots are solutions to the equation \eqref{eq:eomannulus} for $d=2$ and $z_F=1$. Note that for larger radii $\rho_1$ and $\rho_2$ the resulting minimal surface goes deeper into the IR ($z\to\infty$) than it is the case for smaller ones. Note that the solution is not a function, therefore we first generate the blue curve and afterwards search for the matching red one.}
\label{annulus}
\end{figure}

\begin{figure}
\centering
\vspace{-1cm}
\hspace{-0.5cm}
\includegraphics[angle=0,width=0.45\textwidth]{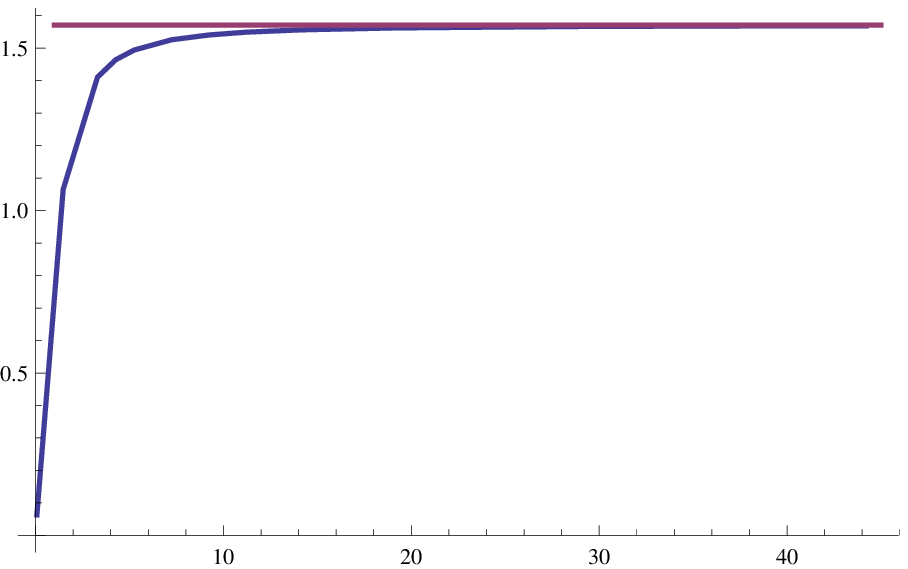}\put(-200,135){$(\Delta \rho)_\text{max}$}\put(-55,-10){$(\rho_2+\rho_1)/2$}
\hspace{-0cm}
\includegraphics[angle=0,width=0.45\textwidth]{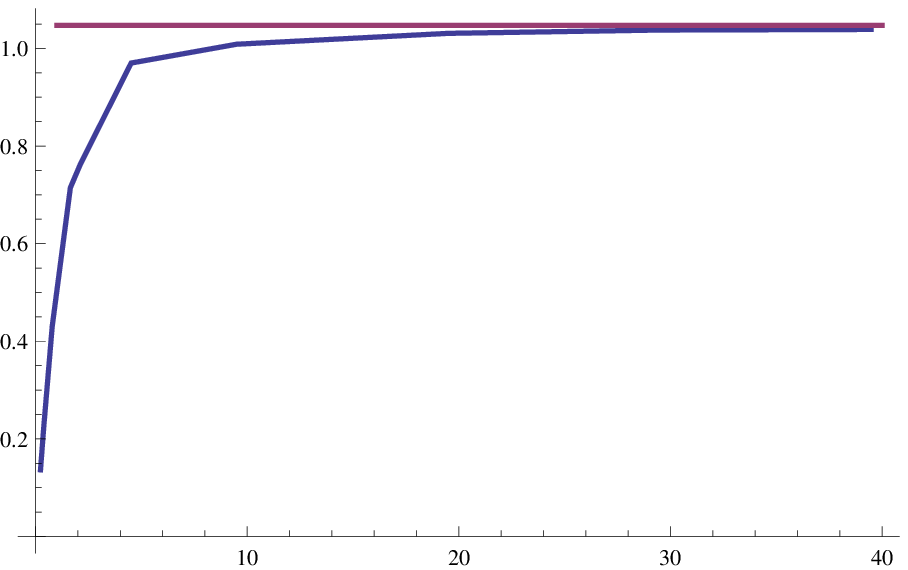}\put(-200,135){$(\Delta \rho)_\text{max}$}\put(-55,-10){$(\rho_2+\rho_1)/2$}
\vspace{0.5cm}\\
\caption{\small We plot the maximal difference between the radii of the connected solution $(\Delta \rho)_\text{max}$ versus the middle radius $(\rho_2+\rho_1)/2$, for $d=2$ (left) and $d=3$ (right). It is apparent that $(\Delta \rho)_\text{max}\to z_F \pi/d$ (red line), with $z_F=1$, for larger values of the radii $\rho_i$.}
\label{fig:deltarhocrit}
\end{figure}

We show generic results for the entanglement entropy for $d=2$ and $d=3$ in figures \ref{Afanulus2} and \ref{Afanulus3}. There we plot $A_\text{finite}$ versus the difference of the radii $\Delta \rho = \rho_2-\rho_1$. We find two connected solutions (deformed annulus, see figure \ref{annulus}) for values of $\Delta \rho \leq (\Delta \rho)_\text{max}$ and one disconnected solution (two concentric balls) for all values of $\Delta \rho$. Note that for each value of $\Delta \rho$ the preferred solution is the one with smaller value of $A_\text{finite}$. In the $d=2$ case, at a value $(\Delta \rho)_\text{crit}$ we find a first order transition from the preferred connected to the disconnected solution for small values of the radii and a second order transition for larger ones. For $d=3$ we find a different behavior: In that case we cannot find a transition for values $\rho_1,\, \rho_2<\mathcal{O}(1)$ (the exact value is hard to find, due to difficult numerical computations), only for large radii we find a second order transition. This behavior is very similar to the strip case discussed in section \ref{sec:strip}, where there also only exists a connected solutions for $l\leq l_\text{crit}$, however, there the transition is second order opposed to the case at hand. The analogy goes further: increasing the values of the radii $\rho_1,\, \rho_2$ leads to $(\Delta \rho)_\text{crit}\to \pi/d$ (c.f.~eq.~\eqref{eq:lcrit} with $z_F=1$). We are not able to check this limit analytically, however our results using numerical methods are in very good agreement with above statement for $d=2$ and $d=3$ (see figure \ref{fig:deltarhocrit}). Looking closer at this limit in $d=2$, we see the swallow tail becomes smaller turning into a second order transition (see right part of fig.~\ref{Afanulus2}). From this behavior we deduce that the annulus tends towards the strip solution for large radii. The other limit, however, where we aim at approximating a sphere, does not work entirely as expected, since for each given pair of radii of the annulus solution, we always find a maximal difference $(\Delta \rho)_\text{max}$ between both which is smaller than outer radius $\rho_2$. Therefore we can at most approximate two concentric spheres, but never one sphere alone. Even this is not always possible, as the small radii $d=3$ case described above shows. Nevertheless, the similarity in most of the parameter space to the behavior seen in confining geometries is astonishing (see \cite{Klebanov:2007ws,Pakman:2008ui}). It would be interesting to understand if there is a common origin to this resemblance.

Finally in the annulus as well as in the strip case $l_\text{crit}=\pi z_F/d$ plays an important governing the phase transition, however, to our knowledge, there is no known dual interpretation of this value. This would be interesting to study. Since it is possible to embed the solution described in section \ref{sec:back} into string theory, at least for the $\eta=1$ case (see~\cite{Kulaxizi:2012gy}), in principle it should be possible to compute the entanglement entropy in the dual theory, although probably this is not feasible from a technical point of view.

\section{Summary and discussion}
We considered the holographic entanglement entropy of semi-local quantum liquids, whose gravity dual in $(d+2)$-dimensions is described by a metric which is conformal to $AdS_{2}\times\mathbb{R}^{d}$ in the IR. The near-horizon geometry is an exact solution of the Einstein-Maxwell-Dilaton theory and its thermodynamics may be investigated, at least qualitatively, by dimensional analysis and scaling arguments. We calculated the holographic entanglement entropy in the IR geometry for both a strip and a sphere. In this geometry, the width of the strip at the boundary $l$ is always constant and the disconnected surface dominates when $l>l_{\rm crit}$. The phase transition between the strip and the two disconnected slabs becomes apparent in the full geometry, which is asymptotically AdS and possesses semi-locality in the IR. When the value of the turning point $z_{\ast}$ is small, i.e. the boundary separation length $l$ is small and the surface probes the UV part of the geometry, $l$ is a smooth function of $z_{\ast}$ and the connected surface dominates. When $z_{\ast}$ is sufficiently large, $l$ approaches the critical value $l_{\rm crit}$ and the disconnected surface dominates. However, such a phase transition is not observed for the spherical entangling region, neither in the IR nor in the full solution. In this case we find for the full solution that $z(r=0)$ (turning point of the spherical solution) grows exponentially with $R$. This is in agreement with equation (A5) of~\cite{Liu:2013una} ($z_t$ in that paper corresponds to $z(r = 0)$ in our case). In addition, the holographic entanglement entropy may be calculated analytically in the IR for the spherical case, which matches the numerical results very well.

In order to interpret the behavior of holographic entanglement entropy with different entangling regions, we also considered the annulus case as an interpolating geometry between the sphere and the strip, following~\cite{Kulaxizi:2012gy}. For annulus type solutions we have to rely on numerics, therefore we restrict ourselves to the $d=2$ and $d=3$ case. The common features in both dimensions are that we find two connected solutions, with one being preferred over the other, i.e.~it has lower entanglement entropy. Furthermore for certain values of $\Delta \rho$, the difference between the outer and inner radius of the annulus at the boundary, we see a transition from the preferred connected solution to a disconnected solution (two concentric spheres). In the $d=2$ case there always seems to be a transition,which is first order for small values of the outer radius and becomes a second order transition for increasing values of the radii. In the $d=3$ case for small values of the radii we cannot find any transition at all. Increasing the values of $\rho_i$ leads to a second order transition. There is no indication of a first order transition in $d=3$. Finally there is a maximal value of $\Delta \rho = (\Delta \rho)_\text{max}$ for a connected solution which in the limit of large radii $\rho_i$ tends towards the value of the critical length of the strip $l_\text{crit}$. Therefore for large radii we approximate the strip. In the other limit of small radii, our solution approaches the case of two concentric spheres, since $\Delta \rho < \rho_2$ with $\rho_2$ being the outer radius, and not to a single sphere with vanishing inner radius. 
 
The behavior of HEE in our background is in parts similar to that in confining backgrounds~\cite{Klebanov:2007ws, Pakman:2008ui}. It was observed in~\cite{Klebanov:2007ws} that when the entangling region is a strip in confining backgrounds, there exist two different types of minimal surfaces. The connected surface dominates when the boundary separation length $l$ is smaller than a critical value $l_{\rm crit}$ while the disconnected one dominates when $l>l_{\rm crit}$. This is very similar to what we see in our strip case. However, for spherical entangling regions in confining geometries a phase transition was also observed in~\cite{Pakman:2008ui} opposed to what we get.

From the discussions above we see that the spherical solution seems to be special. This may be due to a missing scale, as proposed in~\cite{Kulaxizi:2012gy}, which in the strip case comes from the anisotropy of the system and in the annulus case corresponds to the middle radius $(\rho_1+\rho_2)/2$.

A further interpretation for this behavior is provided in~\cite{Liu:2013una}, where the renormalized entanglement entropy near an IR fixed point is extensively studied.
The background metric reads
\begin{equation}
ds^{2}=\frac{L^{2}}{z^{2}}\left(-dt^{2}+d\vec{x}^{2}+\frac{dz^{2}}{f(z)}\right),
\end{equation}
where $f(z)$ can approach either a constant or a power law function $az^{n}, ~a>0,~z>0$ in the IR $z\rightarrow\infty$. Clearly our case falls into the class with $n=2$. Therefore we also find the exponential behavior of the turning point of the minimal surface corresponding to the spherical entangling region described above. Moreover, it was observed in~\cite{Liu:2013una} that the geometry with $n\geq2$ describes a gapped phase while for $n=2$ the system has a continuous spectrum above the gap $\Delta=d/2$. They argued that the presence of a continuum above a gap may be responsible for the peculiar behavior of the HEE. Note that in our case the critical value of the boundary separation length can be rewritten as
\begin{equation}
l_{\rm crit}=\frac{\pi z_F }{d}=\frac{2\pi z_F}{\Delta},
\end{equation}
which may provide certain physical interpretation of $l_{\rm crit}$, relating it to the gap. It would be interesting to explore the connections between the behavior of HEE and the energy gap of the system in future work.

\vspace{1cm}
\bigskip \goodbreak \centerline{\bf Acknowledgments}
\noindent  DWP thanks Andrei Parnachev for sharing his unpublished notes and helpful discussions. We would like to thank Mariano Cadoni, Elias Kiritsis, Manuela Kulaxizi and Koenraad Schalm for interesting discussions and comments.
DWP is supported by Alexander von Humboldt Foundation.

\newpage

\end{document}